\documentclass[aps,prd,groupedaddress,nofootinbib,superscriptaddress,preprint]{revtex4-1}
\usepackage{amsfonts}
\usepackage{amssymb}
\usepackage{amsmath,color}
\usepackage{epsfig}
\usepackage{graphicx,epsf,epsfig}
\usepackage{bbm}
\def\slc#1{\setbox0=\hbox{$#1$}           
    \dimen0=\wd0                                 
    \setbox1=\hbox{/} \dimen1=\wd1               
    \ifdim\dimen0>\dimen1                        
       \rlap{\hbox to \dimen0{\hfil/\hfil}}      
       #1                                        
    \else                                        
       \rlap{\hbox to \dimen1{\hfil$#1$\hfil}}   
       /                                         
    \fi}

\newcommand{\eg}{e.g.}
\newcommand{\ie}{i.e.}

\begin{document}

\preprint{MPP-2011-2}

\title{Renormalization Group Running of the Neutrino Mass Operator in Extra Dimensions}
\author{Mattias Blennow}
\email{blennow@mppmu.mpg.de}

\affiliation{Max-Planck-Institut f{\"u}r Physik
(Werner-Heisenberg-Institut), F{\"o}hringer Ring 6, 80805
M{\"u}nchen, Germany}

\author{Henrik Melb{\'e}us}
\email{melbeus@kth.se}

\author{Tommy Ohlsson}
\email{tommy@theophys.kth.se}

\affiliation{Department of Theoretical Physics, School of
Engineering Sciences, Royal Institute of Technology (KTH) --
AlbaNova University Center, Roslagstullsbacken 21, 106 91 Stockholm,
Sweden}

\author{He Zhang}
\email{he.zhang@mpi-hd.mpg.de}

\affiliation{Max-Planck-Institut f{\"u}r Kernphysik, Postfach
103980, 69029 Heidelberg, Germany}

\begin{abstract}

We study the renormalization group (RG) running of the neutrino masses
and the leptonic mixing parameters in two different
extra-dimensional models, namely, the Universal Extra Dimensions
(UED) model and a model, where the Standard Model (SM) bosons probe
an extra dimension and the SM fermions are confined to a
four-dimensional brane. In particular, we derive the beta function
for the neutrino mass operator in the UED model. We also rederive
the beta function for the charged-lepton Yukawa coupling, and confirm some of the existing results in the
literature. The generic features of the RG
running of the neutrino parameters within the two models are
analyzed and, in particular, we observe a power-law behavior for the
running. We note that the running of the leptonic mixing angle
$\theta_{12}$ can be sizable, while the running of $\theta_{23}$ and
$\theta_{13}$ is always negligible. In addition, we show that the
tri-bimaximal and the bimaximal mixing patterns at a high-energy
scale are compatible with low-energy experimental data, while a tri-small
mixing pattern is not. Finally, we
perform a numerical scan over the low-energy parameter space to
infer the high-energy distribution of the parameters. Using this
scan, we also demonstrate how the high-energy $\theta_{12}$ is
correlated with the smallest neutrino mass and the Majorana phases.

\end{abstract}

\maketitle


\section{introduction}
\label{introduction}

With the recent start-up of the Large Hadron Collider (LHC),
experimental physics has started the search for the domain beyond
the current Standard Model (SM) of particle physics. In addition to
searching for the Higgs boson, the LHC will supply us with important
information about the nature of physics above the TeV energy scale.
Among the most popular high-energy extensions of the SM are
extra-dimensional models. The idea that spacetime could have more
than four dimensions was first proposed by Theodore Kaluza
\cite{Kaluza:1921tu} and Oskar Klein \cite{Klein:1926tv} at the
beginning of the twentieth century. In the 1980's, this idea gained
popularity through the emergence of string theory, and at the end of
the 1990's, several extra-dimensional models, which could
potentially be detected at the next generation of high-energy
experiments, were proposed
\cite{Antoniadis:1998ig,ArkaniHamed:1998rs,ArkaniHamed:1998nn,Randall:1999vf,Randall:1999ee,Appelquist:2000nn}.
These models are mainly motivated by the fact that they could
provide solutions to different problems in the SM, such as the
hierarchy problem and the lack of a good particle dark matter
candidate.

An interesting aspect of extra dimensions is their impact on the
RG running of physical parameters. It has
been shown that the Kaluza--Klein (KK) towers give rise to an
effective power-law running of the parameters for energies above the
first KK level~\cite{Dienes:1998vg}. Hence, extra dimensions may
increase the RG running dramatically, resulting in large effects at
relatively low energy scales.

One of the rare examples of experimental evidence for physics beyond
the SM comes from neutrino physics. The observation of neutrino
oscillations strongly indicates that neutrinos are massive and that
lepton flavors are mixed. Since it is not possible to describe
neutrino masses within a renormalizable framework using the SM
particle content only, neutrino oscillations imply new physics
beyond the SM. The fact that the neutrino masses are bounded to be
unnaturally small in comparison to the other SM fermion masses,
together with the possibility that neutrinos could be their own
antiparticles, has given rise to extensive model building within the
neutrino sector. For example, small neutrino masses could naturally
be generated through the so-called seesaw mechanisms
\cite{Minkowski:1977sc,Yanagida:1979as,Mohapatra:1979ia,GellMann:1980vs},
where neutrinos couple to new degrees of freedom. Small neutrino
masses could also be generated in certain extra-dimensional models
\cite{Frere:2003hn,Haba:2009sd,Blennow:2010zu,Saito:2010xj}, which
have the advantage of potentially being observable at the LHC.

Since the neutrino parameters are measured in low-energy scale
experiments, the RG running effects should be taken into account
properly in studying neutrino mass models at high-energy
scales~\cite{Chankowski:1993tx,Babu:1993qv,Antusch:2001ck,Antusch:2001vn,Chao:2006ye,Schmidt:2007nq,Chakrabortty:2008zh,Bergstrom:2010qb}.
Therefore, in this paper, we study the RG running of the neutrino
parameters in extra-dimensional models. We employ an effective
description for neutrino masses in terms of the dimension-five
Weinberg operator, and investigate the RG running of the neutrino
parameters in two different extra-dimensional models. In particular,
we derive the beta function for the Weinberg operator in the
Universal Extra Dimensions (UED) model \cite{Appelquist:2000nn},
which has previously not been performed in the literature. In order
to determine which high-energy values that are consistent with
current experimental data within the models, we use a Markov Chain
Monte Carlo to infer the favored high-energy parameter values from
the low-energy parameter bounds using our analytical results.

The rest of the work is organized as follows: In
Sec.~\ref{sec:EDRGE}, we describe the models that we have studied,
and present the beta functions for the neutrino mass operator in
both models. In Sec.~\ref{sec:nuRunning}, we give the RGEs for the
neutrino masses and the leptonic mixing parameters, which are
obtained from the beta function for the neutrino mass operator.
Then, in Sec.~\ref{sec:NumericalAnalysis}, we show our numerical
results for the RG running. Finally, in Sec.~\ref{sec:Summary}, we
summarize our work and state our conclusions. In addition, in
Appendix \ref{sec:appendix}, we give the full set of RGEs that we
have used.

\section{RGEs in extra-dimensional theories}\label{sec:EDRGE}

In order to illustrate the features of RGEs in extra-dimensional
theories, we consider two representative models. First, we
investigate the UED model, in which all the SM fields probe the
extra spatial dimensions. Then, we study a model, in which the SM
bosons propagate in the bulk, while the SM fermions are localized to a
four-dimensional brane \cite{Dienes:1998vg}.

In general, the KK towers in an extra-dimensional theory are
expected to be cut off at some energy scale $\Lambda$ in order to
keep the theory renormalizable, with the nature of the cutoff scale
depending on the specific ultraviolet (UV) completion of the model.
For the purpose of illustration, we take $\Lambda = 50~{\rm TeV}$
for both models in this work.

\subsection{Model~I: Universal Extra Dimensions}

The UED model is constructed by promoting all the SM fields to a
higher-dimensional flat spacetime, and hence, all the SM particles
acquire towers of KK modes. Here, we consider the simplest case with
only a single extra spatial dimension, which is assumed to be
compactified on an $S^1/\mathbb{Z}_2$ orbifold with radius $R$. In
this framework, KK parity, which is defined as $(-1)^n$ for the
$n{\rm th}$ KK level, is conserved after compactification. The mass
scale of the first excited KK level, given by $R^{-1}$, is bounded
to be larger than approximately 300 GeV \cite{Hooper:2007qk}. In
this work, we use the value $R^{-1} = 1~{\rm TeV}$.

In addition to the operators that are renormalizable at the level of
the SM, we introduce the dimension-five Weinberg operator
responsible for neutrino masses
\begin{equation}\label{eq:kappa}
-{\cal L}_\nu = \frac{1}{2} \left(\overline{L}
\phi\right) \hat \kappa \left(\phi^T L^c\right) + {\rm h.c.}\,
,
\end{equation}
where $L$ denotes the lepton doublet fields, $\phi$
denotes the Higgs doublet, and $\hat \kappa$ is a matrix in
flavor space. After electroweak
symmetry breaking, the neutrino mass matrix is obtained as
\begin{eqnarray}\label{eq:mnu}
m_\nu \equiv \kappa v^2 \, ,
\end{eqnarray}
where $v \simeq 174 ~ {\rm GeV}$ is the vacuum expectation value of
the Higgs field and $\kappa \equiv \hat \kappa / \pi R$. The neutrino mass operator defined in
Eq.~\eqref{eq:kappa} can be realized through certain
extra-dimensional seesaw mechanisms at the cutoff scale
$\Lambda$~\cite{Blennow:2010zu}, as well as through the standard
seesaw mechanisms. In the language of an effective theory,
Eq.~\eqref{eq:kappa} is essentially the same for the different
seesaw models.

In a five-dimensional spacetime, there are no chiral fermions. In
order to reproduce the phenomenology of the SM for energies below
$R^{-1}$, a five-dimensional Dirac fermion has to be introduced for
each chiral fermion in the SM, and the KK expansions of these Dirac
fermions are chosen in such a way that chiral fermions are obtained
at the zero-mode level. Hence, the number of degrees of freedom in
the fermion sector is doubled at the excited KK levels in comparison
to the SM. In addition, each SM gauge field has a fifth component,
which appears as a real scalar from the four-dimensional point of
view. Again, the zero-modes of these scalars can be removed by
suitable choices of KK expansions, but they appear at the excited KK
levels. In particular, compared to the SM-like couplings, additional
vertices involving SM fermions and the fifth components of gauge
fields appear.

In the UED model, the beta function for the neutrino mass operator
can be written as
\begin{eqnarray}\label{eq:RGEUED}
16\pi^2 \frac{{\rm d}\kappa}{{\rm d}\ln \mu}  = \beta^{\rm
SM}_\kappa + \beta^{\rm UED}_\kappa \, ,
\end{eqnarray}
where $\beta^{\rm SM}_\kappa$ denotes the SM beta
function~\cite{Babu:1993qv,Chankowski:1993tx,Antusch:2001vn}
\begin{eqnarray}\label{eq:betaSM}
\beta_\kappa^{\rm SM} = -\frac{3}{2} \kappa (Y_\ell^\dagger Y_\ell)
- \frac{3}{2} (Y_\ell^\dagger Y_\ell)^T \kappa - \left( 3 g_2^2 - 2T
- \lambda \right) \kappa,
\end{eqnarray}
where
\begin{eqnarray}\label{eq:T}
T  = {\rm tr} \left(3 Y^\dagger_u Y_u + 3 Y^\dagger_d Y_d +
Y^\dagger_\ell Y_\ell \right) \, ,
\end{eqnarray}
with $Y_f$ (for $f=u,d,\ell$) denoting the Yukawa coupling matrices
of the up-type quarks, down-type quarks, and charged leptons,
respectively. Here, $g_i$ are the gauge couplings and $\lambda$
denotes the Higgs self-coupling constant. The second term on the
right-hand side of Eq.~\eqref{eq:RGEUED}, $\beta^{\rm UED}_\kappa$,
comes from the contributions of the excited KK modes. As mentioned
above, the KK spectrum at the excited levels differs from the SM,
and this is reflected in the contributions of the KK modes to the
beta function. We have calculated $\beta^{\rm UED}_\kappa$ with the
following result
\begin{eqnarray}\label{eq:betaUED}
\beta^{\rm UED}_\kappa  &= & s \left[-\frac{3}{2} \kappa \left(
Y^\dagger_\ell Y_\ell \right)- \frac{3}{2} \left( Y^\dagger_\ell
Y_\ell \right)^T \kappa -\left( \frac{1}{4} g^2_1 + \frac{11}{4}
g^2_2 - 4 T - \lambda\right) \kappa \right] \, ,
\end{eqnarray}
where $s = \lfloor \mu/\mu_0 \rfloor$ counts the number of KK levels contributing to
the beta function for a given energy $\mu$. Here, $\mu_0 = R^{-1} = 1 \, {\rm TeV}$. For large $\mu / \mu_0$, $s$ is well-approximated by the continuous expression $\mu/\mu_0$. In evaluating physical
parameters from lower to higher energy scales, new KK excitations
enter the theory at each KK threshold, giving additional quantum
corrections, and hence, the coefficients of the beta functions are
modified depending on how many KK modes that are excited. The result
is that the RG running of $\kappa$ follows a power-law behavior,
controlled by $s$. This is a distinctive feature of RG running in
extra-dimensional theories. Compared to the RG running behavior in the
SM, where the coefficients of the beta functions are nearly
constant, the power-law behavior results in a significant boost in
the RG running, which could possibly be tested at near-future
experiments.

\begin{figure*}
\begin{center}
\includegraphics[width=.75\textwidth]{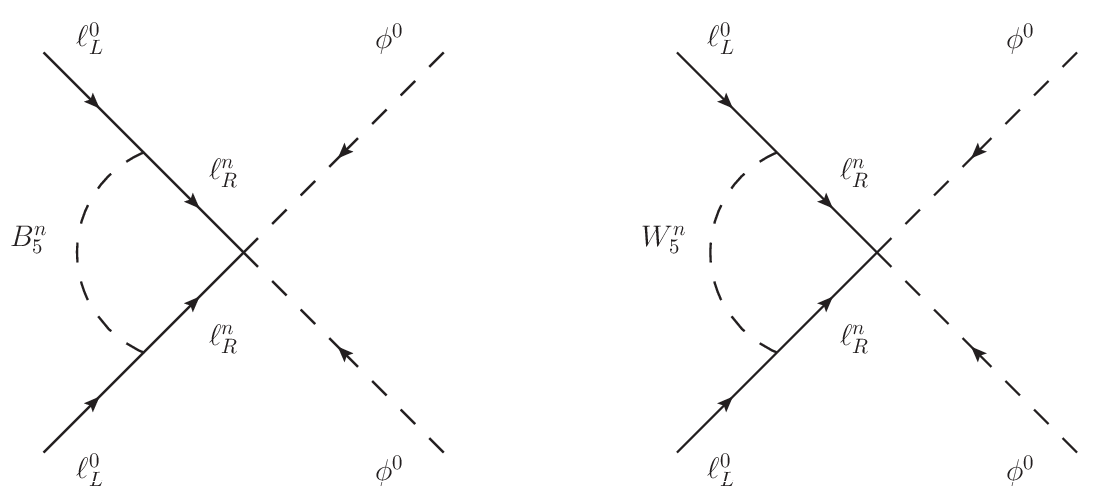}
\includegraphics[width=.75\textwidth]{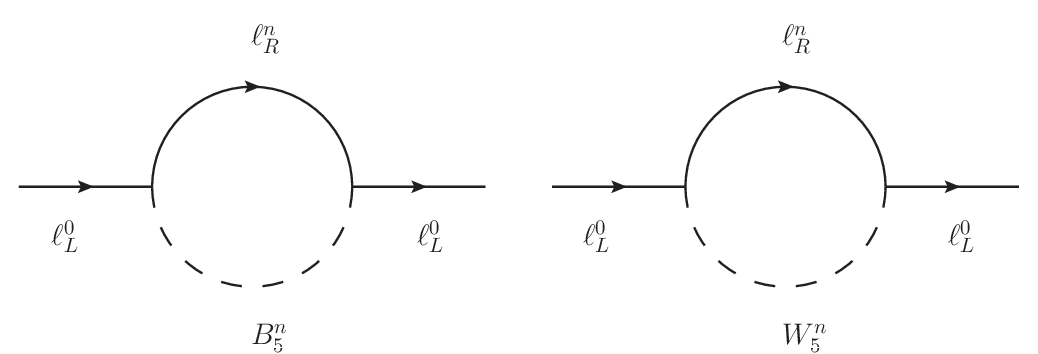}
\caption{\label{fig:RGE-UED-kappa} The Feynman diagrams including the fifth components of the electroweak gauge
bosons that contribute to the beta function for $\kappa$.}
\end{center}
\end{figure*}

The differences in the coefficients for the gauge couplings between
Eqs.~\eqref{eq:betaSM} and~\eqref{eq:betaUED} are due to the additional
Feynman diagrams involving the fifth components of the electroweak gauge
bosons shown in Fig.~\ref{fig:RGE-UED-kappa}. The additional factor of $2$ in the coefficient for $T$ is
due to the fact that the chiral fermions are replaced by Dirac
fermions at each excited KK level.

The beta functions for the Yukawa couplings $Y_f$ as well as the
gauge couplings are listed in Appendix \ref{sec:appendix}. Note that
the beta functions for the Yukawa couplings differ between
Refs.~\cite{Bhattacharyya:2006ym} and \cite{Cornell:2010sz}. We have
confirmed the computations in Ref.~\cite{Cornell:2010sz}, namely,
\begin{eqnarray}\label{eq:Yl}
\beta^{\rm UED}_{Y_\ell} = s Y_\ell \left[ \frac{3}{2}
Y^\dagger_\ell Y_\ell + 2 T - \frac{33}{8} g^2_1 - \frac{15}{8}
g^2_2 \right] \, .
\end{eqnarray}

\subsection{Model~II: Fermions on the brane}

In this model, the SM fermions are confined to a four-dimensional
brane, while all the SM bosons probe the bulk \cite{Dienes:1998vg}.
Again, we assume the extra dimension to be compactified on an
$S^1/\mathbb{Z}_2$ orbifold with radius $R$. Present collider bounds
allow the masses of the lowest KK excitations to be
as low as 500 GeV~\cite{Dienes:1998vg}. As for model~I, we take
$\mu_0 = R^{-1} = 1~{\rm TeV}$ in this work. Thus, the heavy KK
modes of the SM particles could be accessible at forthcoming
collider experiments, and the unification of gauge couplings could
also be achieved at a low-energy
scale~\cite{Antoniadis:1990ew,Dienes:1998vh}.

Again, we write the beta function for $\kappa$ as
\begin{eqnarray}
16\pi^2 \frac{{\rm d}\kappa}{{\rm d}\ln \mu}  = \beta^{\rm
SM}_\kappa + \beta^{\rm II}_\kappa \, ,
\end{eqnarray}
where the SM beta function $\beta^{\rm SM}_\kappa$ is given in
Eq.~\eqref{eq:betaSM}. The contribution from the excited KK modes
reads \cite{Dienes:1998vg,Abel:1998wa,Bhattacharyya:2002nc}
\begin{eqnarray}\label{eq:RGE}
\beta^{\rm II}_\kappa = 2s \left[ -\frac{3}{2} \kappa
(Y_\ell^\dagger Y_\ell) - \frac{3}{2} (Y_\ell^\dagger Y_\ell)^T
\kappa - \left( 3 g_2^2 - \lambda \right) \kappa \right].
\end{eqnarray}
In comparison to the SM beta function, the term proportional to $T$
is missing here. This term comes from the fermion loop contributions
to the self-energy of the Higgs boson, and since the fermions have
no KK excitations in this model, there is no such contribution for
$n>0$. Also, the remaining terms in Eq.~\eqref{eq:RGE} are larger by
a factor of 2, which is due to a rescaling of all interactions
involving KK excitations by a factor $\sqrt{2}$, coming from the
canonical normalization of all KK modes \cite{Rizzo:1999br}. Note
that this factor does not exist in model~I.

\section{Running neutrino parameters}\label{sec:nuRunning}

We proceed our discussion to the RG running of the
neutrino parameters. In general, one can choose to work
in a basis where the charged lepton Yukawa coupling matrix $Y_\ell$
is diagonal, \ie, $Y_\ell ={\rm diag}(y_e,y_\mu,y_\tau)$. In this
basis, the leptonic mixing
matrix~\cite{Maki:1962mu,Pontecorvo:1967fh} stems from the
diagonalization of the neutrino mass matrix, \ie,
\begin{eqnarray}\label{eq:U}
U^\dagger m_\nu U^*= D_\nu \equiv {\rm diag} (m_1,m_2,m_3) \, ,
\end{eqnarray}
with $m_i$ being the neutrino masses. Inserting Eq.~\eqref{eq:U}
into the beta function for $\kappa$, one obtains the evolution for
the leptonic mixing matrix due to the KK modes as
\begin{eqnarray}\label{eq:RGEU}
\frac{{\rm d}U}{{\rm d}t}\equiv\dot{U} = 2s U X \, ,
\end{eqnarray}
with $t= \ln \mu / 16\pi^2$, and
\begin{eqnarray}\label{eq:ReT} {\rm
Re}{X}_{ij} & = & -\frac{3}{2} \eta \zeta_{ij} {\rm
Re}\left(U^\dagger Y^\dagger_\ell Y_\ell U
\right)_{ij} \, , \\
{\rm Im}{X}_{ij} & = & -\frac{3}{2} \eta \zeta^{-1}_{ij} {\rm
Im}\left(U^\dagger Y^\dagger_\ell Y_\ell U \right)_{ij} \, ,
\label{eq:ImT}
\end{eqnarray}
where $\eta=1$ in model~I, and $\eta=2$ in model~II, and the indices
$i$ and $j$ run over 1, 2, and 3. The factors $\zeta_{ij}$ are
defined as $\zeta_{ij}=(m_j+m_i)/(m_j-m_i)$. Note that the charged
lepton mass spectrum is strongly hierarchical, \ie, $m_e \ll m_\mu
\ll m_\tau$, which allows us to make a reasonable approximation by
ignoring the electron and muon Yukawa couplings in
Eqs.~\eqref{eq:ReT} and \eqref{eq:ImT}. In what follows, we will
assume $Y_\ell = {\rm diag} (0,0,y_\tau)$ for simplicity.
Furthermore, the $\zeta$ factors play a key role in the RG running
of the neutrino mixing angles, since in the case of a nearly
degenerate neutrino mass spectrum, \ie, $m_1 \simeq m_2 \simeq m_3$,
$\zeta_{ij} \gg 1$, and therefore, the RG running effects will be
enhanced dramatically. If the neutrino mass spectrum is
hierarchical, \eg, $m_1 \ll m_2 \ll m_3$, $\zeta_{ij} \simeq 1$
holds and the RG running effects on the leptonic mixing matrix are
not observable. In the following analytical analysis, we will assume
a nearly degenerate neutrino mass spectrum.

In order to figure out the RG running behaviors of the leptonic mixing
parameters, we employ the standard parametrization, in which $U$ is
parametrized by three mixing angles and three CP-violating phases as
\begin{eqnarray}\label{eq:para}
U & = & \left( \begin{matrix}c_{12} c_{13} & s_{12} c_{13} & s_{13}
e^{-{\rm i}\delta} \cr -s_{12} c_{23}-c_{12} s_{23} s_{13} e^{{\rm
i} \delta} & c_{12} c_{23}-s_{12} s_{23} s_{13} e^{{\rm i} \delta} &
s_{23} c_{13} \cr s_{12} s_{23}-c_{12} c_{23} s_{13} e^{{\rm i}
\delta} & -c_{12} s_{23}-s_{12} c_{23} s_{13} e^{{\rm i} \delta} &
c_{23} c_{13}\end{matrix} \right) \left(
\begin{matrix} e^{{\rm i}\rho} & & \cr  & e^{{\rm i}\sigma} & \cr & &
1 \end{matrix} \right) \ ,
\end{eqnarray}
with $c_{ij} \equiv \cos \theta_{ij}$ and $s_{ij} \equiv \sin
\theta_{ij}$ ($ij=12$, $13$, $23$). Combining Eqs.~\eqref{eq:RGEU} and
\eqref{eq:para}, we arrive at the running for the leptonic mixing angles
\begin{eqnarray}\label{eq:theta12approx2}
\dot\theta_{12} & \simeq & \frac{3}{2}\eta s \zeta _{12} s_{12}
c_{12} s_{23}^2 c^2_{\rho -\sigma }
y^2_\tau\, , \\
\dot\theta_{23} & \simeq & \frac{3}{2} \eta s \zeta_{13}
s_{23}c_{23} \left(s_{12}^2c_{\rho }^2 +c_{12}^2 c_{\sigma
}^2\right)
y_\tau^2 \, ,  \\
\label{eq:theta13approx}\dot\theta_{13} & \simeq & \frac{3}{2} \eta
s  \zeta_{13} s_{12} c_{12}s_{23} c_{23} \left(c_{\sigma } c_{\delta
+\sigma }-c_{\rho } c_{\delta +\rho }\right)y_\tau^2  \, ,
\end{eqnarray}
where $c_x \equiv \cos x$. Here, we have taken $\zeta_{13} \simeq \zeta_{23}$ and neglected the smallest mixing angle $\theta_{13}$
to a reasonably good approximation. In the limit of four dimensions,
\ie, $s=\eta=1$, the RGEs for $\theta_{ij}$ in the SM are
reproduced~\cite{Antusch:2005gp,Mei:2005qp}. One observes that,
compared to the RG running of the mixing angles in four-dimensional
theories, an additional enhancement factor $s$ enters the RGEs,
which may lead to significant RG running effects. However, as the beta functions for the mixing angles are proportional to the four-dimensional ones, some qualitative features (e.g., the dependences of the running mixing angles on the neutrino mass hierarchy and the Majorana CP-violating phases) remain the same, and were already known in four-dimensional theories~\cite{Chankowski:1993tx,Babu:1993qv,Antusch:2001ck,Antusch:2001vn,Chao:2006ye,Schmidt:2007nq,Chakrabortty:2008zh,Bergstrom:2010qb}.

For the RG running of the leptonic mixing angles, one of the key
features is that $\theta_{12}$ increases with increasing energy
scale, independently of the neutrino mass hierarchy and the CP-violating
phases. Consequently, a small $\theta_{12}$ at the high-energy scale
is disfavored. On the other hand, the tri-bimaximal
($s_{12}=1/\sqrt{3}, ~s_{23}=1/\sqrt{2},
~s_{13}=0$~\cite{Harrison:2002er,Harrison:2002kp,Xing:2002sw}) and
the bimaximal ($s_{12}=s_{23}=1/\sqrt{2}, ~s_{13}=0$) mixing
patterns could be natural candidates for flavor symmetries at some
high-energy scale. As for $\theta_{23}$ and $\theta_{13}$, the RG
running effects are milder, since their RG running is boosted by
$\zeta_{13}$, which is much smaller than $\zeta_{12}$.

For the sake of completeness, we also give the analytical RGEs for the
neutrino masses
\begin{eqnarray}\label{eq:mass-approx1}
\dot m_1 & \simeq & -\frac{3}{2} \eta s m_1 s_{12}^2 s_{23}^2
y_\tau^2 + m_1 \alpha_{\kappa} \, ,
\\ \label{eq:mass-approx2}
\dot m_2 & \simeq & -\frac{3}{2}  \eta s m_2 s_{23}^2 c_{12}^2
y_\tau^2 + m_2 \alpha_{\kappa} \, ,
\\
\label{eq:mass-approx3}\dot m_3 & \simeq & -\frac{3}{2} \eta s  m_3
c_{23}^2 y_\tau^2+m_3 \alpha_{\kappa} \, ,
\end{eqnarray}
where $\alpha_\kappa$ is flavor universal, and can be found in
Appendix~\ref{sec:appendix}. Similarly to the RG running of the mixing
angles, the power-law factor $s$ enhances the RG corrections.
However, compared to the RGEs for the mixing angles, there is no
enhancement factor $\zeta$. Furthermore, the flavor non-trivial
parts in Eqs.~(\ref{eq:mass-approx1})-(\ref{eq:mass-approx3}) are
suppressed by the charged-lepton Yukawa coupling $y_\tau$ in
comparison to $\alpha_\kappa$. In particular, if one ignores the terms
proportional to $y^2_\tau$, and obtain a very simple form for
Eqs.~(\ref{eq:mass-approx1})-(\ref{eq:mass-approx3}) as $\dot m_i
\simeq m_i \alpha_\kappa$. The solution to this equation is roughly
estimated by
\begin{eqnarray}\label{eq:mass-approx}
\frac{m_i(\Lambda)}{m_i (M_Z)} \simeq
\left(\frac{\Lambda}{M_Z}\right)^{\alpha_\kappa} \, .
\end{eqnarray}
Therefore, the RG running of the neutrino masses is only sensitive to
$\alpha_\kappa$, independently of the neutrino mass spectrum and the
mixing parameters.

In the following numerical analysis, we will mainly concentrate on
the RG corrections to the flavor structure of the leptonic mixing
matrix in the two above models.

\section{Numerical analysis}\label{sec:NumericalAnalysis}

In our numerical computations, we make use of the full sets of RGEs
without any approximations. The input values for the neutrino
parameters and SM observables at the $\mu=M_Z$ scale are taken from
Refs.~\cite{Schwetz:2008er,Xing:2007fb}. In addition, for the Higgs
mass, we use the representative value $m_H=140~{\rm GeV}$. Direct
information on the absolute neutrino mass scale can be derived from
tritium beta decay experiments~\cite{Lobashev:2003kt,Eitel:2005hg},
\ie, $m_{\nu} < 2.3 ~{\rm eV}$ (at $95~\%$ {\rm C.L.}). Indirect
constraints from the CMB data of the WMAP experiment and the large
scale structure surveys also lead to an upper limit on the sum of
neutrino masses. In our numerical studies, we conservatively take
$m_i < 0.5~{\rm eV}$.

\begin{figure*}[t]
\begin{center}\vspace{0.cm}
\includegraphics[width=7cm]{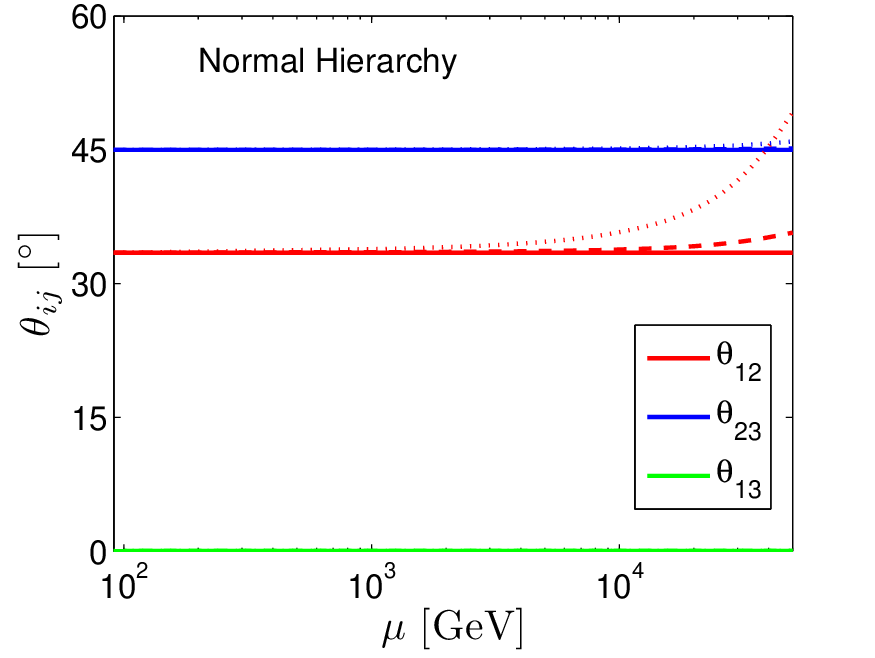}
\includegraphics[width=7cm]{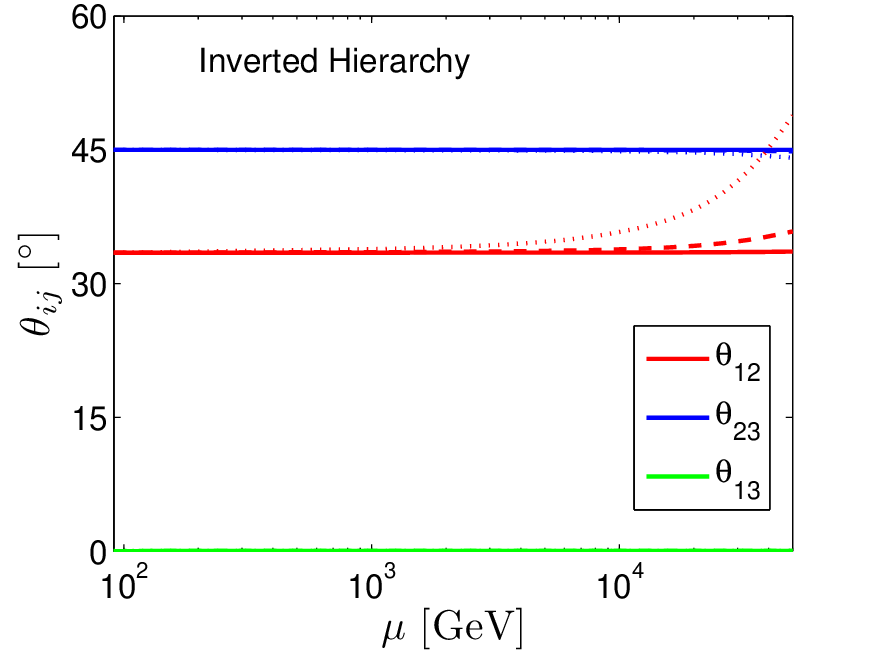}
\includegraphics[width=7cm]{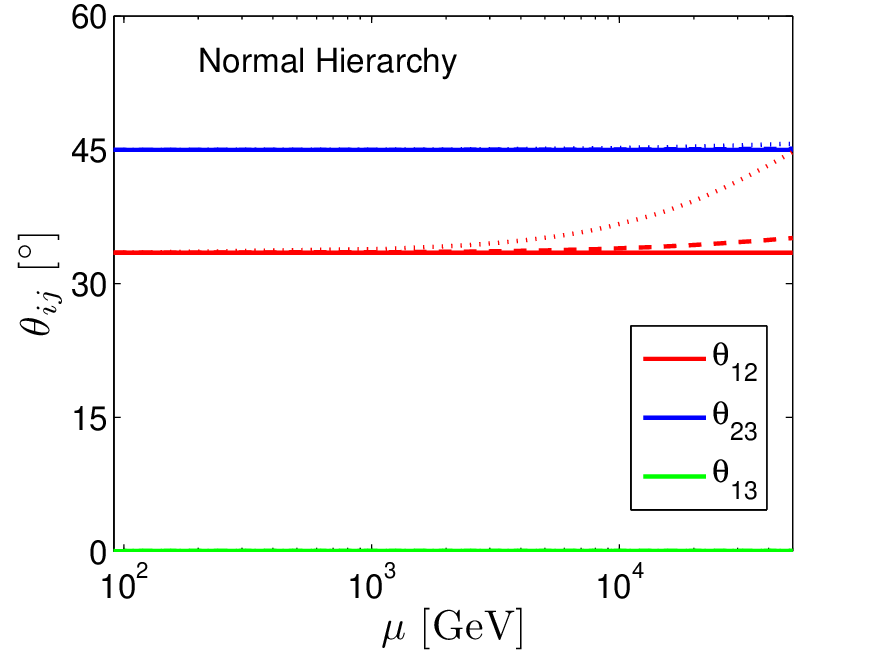}
\includegraphics[width=7cm]{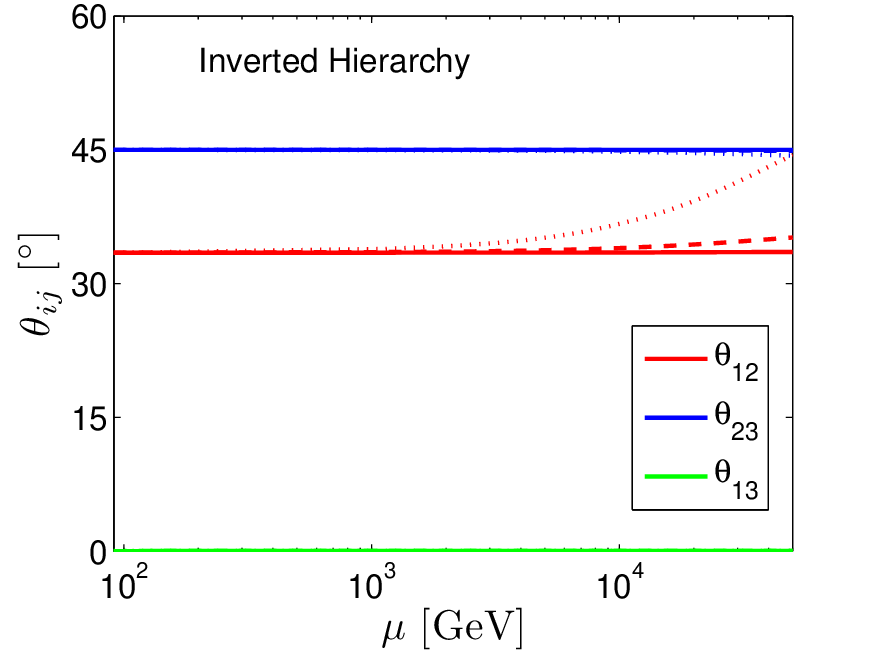}
\caption{\label{fig:fig1} The RG evolution of the three leptonic
mixing angles as functions of the energy scale from $M_Z$ to
$\Lambda$ for the normal mass hierarchy (left column) and the
inverted mass hierarchy (right column) in model~I (upper row) and
model~II (lower row). The solid, dashed, and dotted curves
correspond to the mass of the lightest neutrino to be $0$, $0.2~{\rm
eV}$, and $0.5~{\rm eV}$, respectively. Since there is no strong
experimental evidence for a non-vanishing $\theta_{13}$, we take
$\theta_{13}=0$ as the input value at the $M_Z$ scale. In addition,
all the CP-violating phases are taken to be zero.} \vspace{-0.cm}
\end{center}
\end{figure*}

In Fig.~\ref{fig:fig1}, we show the RG evolution of the neutrino
mixing angles.
In good agreement with our analytical results, $\theta_{12}$ is the
mixing angle, which is the most sensitive to the RG corrections, and
it increases with increasing energy in both the normal hierarchy
($m_1<m_2<m_3$) and inverted hierarchy ($m_3<m_1<m_2$) cases. In
contrast, $\theta_{23}$ and $\theta_{13}$ are rather stable under
the RG running, which reflects the fact that there is no strong
enhancement factor $\zeta_{12}$ in their RGEs. Furthermore, the
running direction of $\theta_{23}$ depends on the mass hierarchy,
namely, an increasing (decreasing) $\theta_{23}$ is obtained in the
case of normal mass hierarchy (inverted mass hierarchy). This
feature can be understood from the sign of $\zeta_{13}$, which is
positive in the normal hierarchy case, and negative in the inverted
hierarchy case.

Note that the RG running effects on the mixing angles in model I are generally larger than those in model II, although the coefficients of the beta functions in model II are twice as large due to $\eta$. This can be understood from the different RG running behavior of $y_\tau$ in the two models. Concretely, in model I, the flavor independent parts of $\beta_{y_\tau}$ contain the trace of charged-fermion Yukawa couplings (c.f., the $T$ term in Eq.~\eqref{eq:alphal}), whereas such a contribution does not exist in model II, due to the absence of fermion KK excitations. Therefore, $y_\tau$ receives larger RG corrections in the UED model and a more sizable $y_\tau$ could naturally be expected at higher scales, which eventually leads to more significant RG corrections to $\theta_{ij}$ in the UED model.

\begin{figure}[t]
\begin{center}\vspace{0.cm}
\includegraphics[width=7cm]{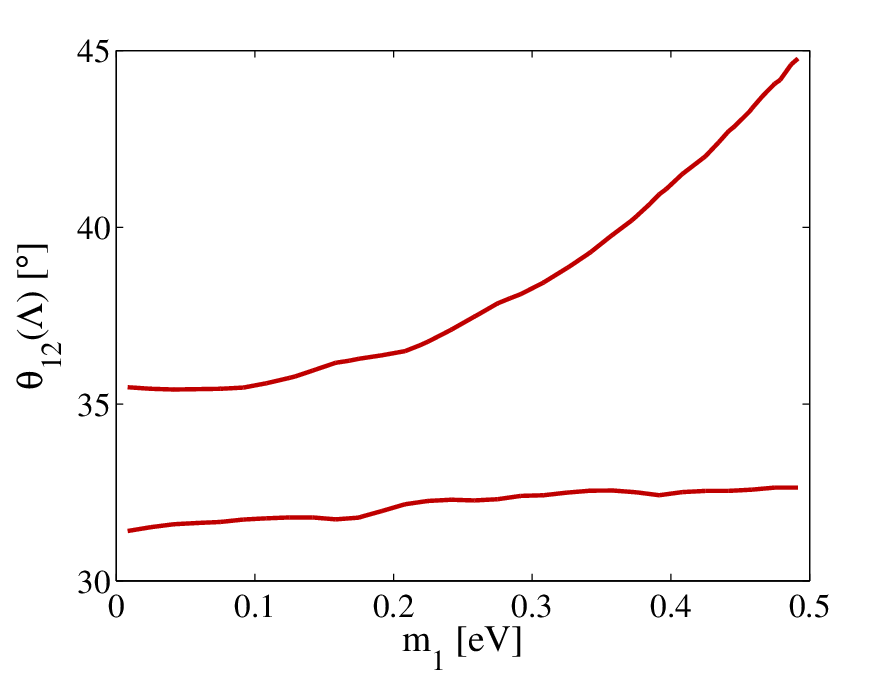}
\caption{\label{fig:fig2} The dependence of $\theta_{12}$ on
$m_1$ at $\Lambda=50~{\rm TeV}$ in model~I. The band between the two
curves corresponds to the allowed parameter space at 90~\% posterior
probability. Here, a normal mass hierarchy is assumed.}
\vspace{-0.cm}
\end{center}
\end{figure}

In Fig.~\ref{fig:fig2}, we present the dependence of the high-energy
$\theta_{12}$ on $m_1$ in the normal mass hierarchy case.
In the numerical calculations, we use the MonteCUBES software
\cite{Blennow:2009pk} as a basis for a Markov Chain Monte Carlo
(MCMC), and generate $5\cdot 10^4$ samples per value of the mass
$m_1$, while the neutrino oscillation parameters have priors
corresponding to present bounds. The band between the two curves
corresponds to the allowed parameter space at 90~\% posterior
probability\footnote{The 90~\% posterior probability region is the
smallest region containing 90~\% of the posterior probability
distribution, \ie, given the priors, there is a 90~\% probability
that the parameters are within this region. It is to Bayesian
statistics what confidence regions is to frequentist statistics.}.
One can observe that remarkable changes in $\theta_{12}$ can be
achieved at the higher energy scales if $m_1$ is sufficiently large.
For a proper choice of $m_1$, the tri-bimaximal or the bi-maximal
mixing pattern can be easily achieved at higher-energy scales.
However, in contrast to the conclusion obtained in
Ref.~\cite{Bhattacharyya:2002nc}, large leptonic mixing angles at
low-energy scales cannot be generated from very small mixing angles
near the cut-off scale. This is due to the naive but
unrealistic two-flavor picture employed in
Ref.~\cite{Bhattacharyya:2002nc}. Explicitly, in the two-flavor
framework, a small mixing angle can be obtained at the cut-off scale
for $m^2/\Delta m^2 \sim 10^{4}$. Thus, when we are working in the
standard three-flavor framework, the RG corrections to $\theta_{13}$
and $\theta_{23}$ are related to $ (m_3+m_2)^2/\Delta m^2_{32}$,
which can be at most at order $10^{2}$ for $m_i < 0.5~{\rm eV}$,
and hence not sufficiently large to result in visible effects. As
for $\theta_{12}$, the RG running is related to the ratio
$(m_2+m_1)^2/\Delta m^2_{21}$, which could be large enough to give significant corrections for a nearly degenerate neutrino mass
spectrum. However, as we have mentioned, the RG running always leads
to a larger value of $\theta_{12}$ at the cut-off scale, independently
of the choices of neutrino mass hierarchy. Therefore, the tri-small
mixing pattern is not compatible with the models under
consideration. We further remark that this feature is generic and
independent of the choices of other physical parameters.

\begin{figure}[t]
\begin{center}\vspace{0.cm}
\includegraphics[width=7.5cm]{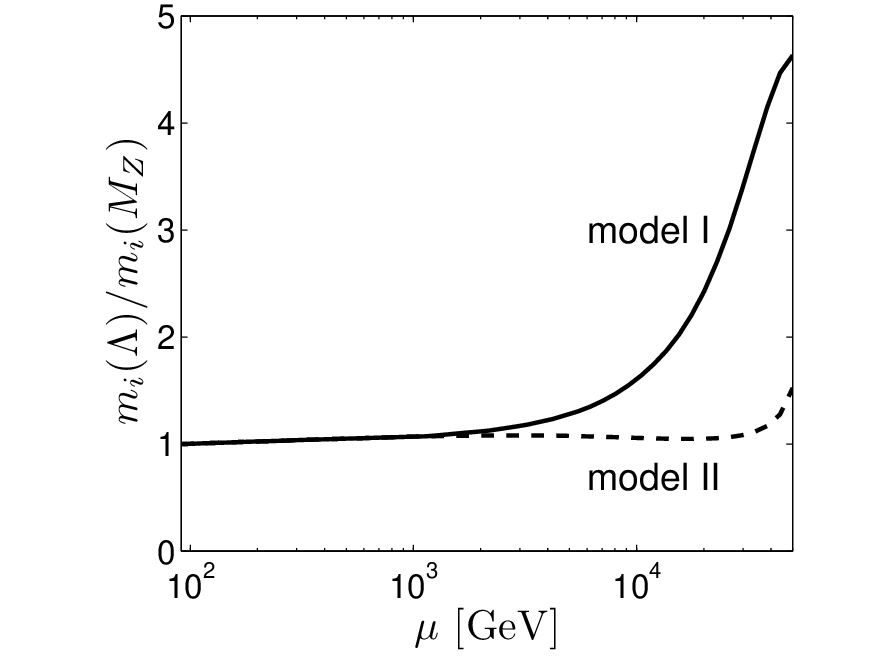}
\caption{\label{fig:figm} The RG evolution of the ratio
$m_i(\Lambda)/m_i(M_Z)$ on $m_1$ for model~I (solid curve) and model
II (dashed curve). Here, we use $m_1 = 0.2~{\rm eV}$ and a normal
mass hierarchy.} \vspace{-0.cm}
\end{center}
\end{figure}

For the sake of completeness, we illustrate the running of the
neutrino masses in Fig.~\ref{fig:figm}.
As expected, the RG running behavior is universal for the three
neutrino masses within a specific model. For model~I, the RG
corrections could increase the neutrino masses dramatically, whereas
in model~II, the RG corrections are not significant for the neutrino
masses. This is consistent with our analytical result that
$\alpha_\kappa$ is larger in model~I, due to the contributions to
$T$ from the top quark Yukawa coupling. We have also checked that
this conclusion does not change for different hierarchies of the
neutrino masses.

\begin{figure}[h]
\begin{center}\vspace{0.cm}
\includegraphics[width=7.5cm]{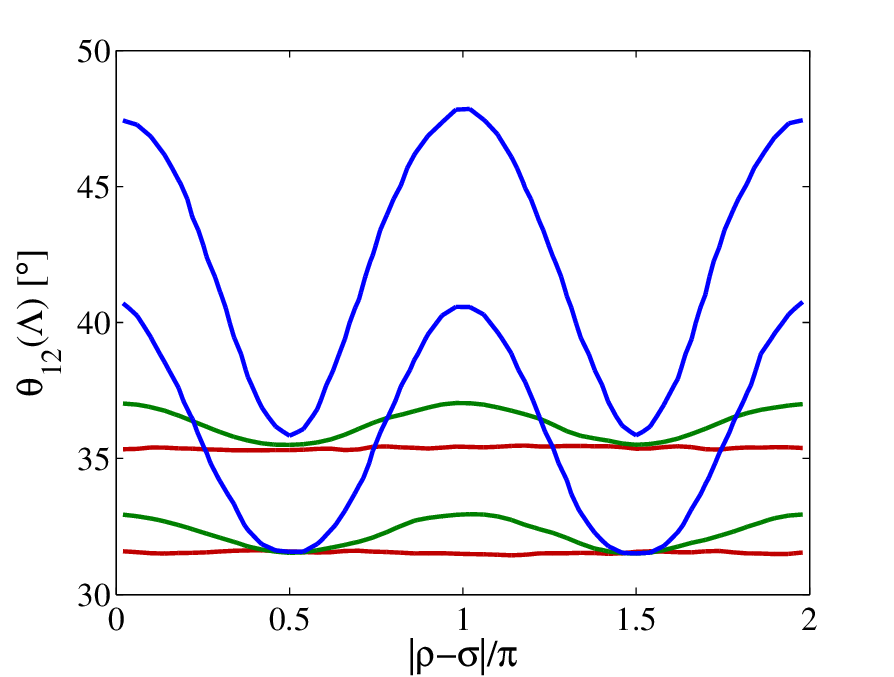}
\includegraphics[width=7.5cm]{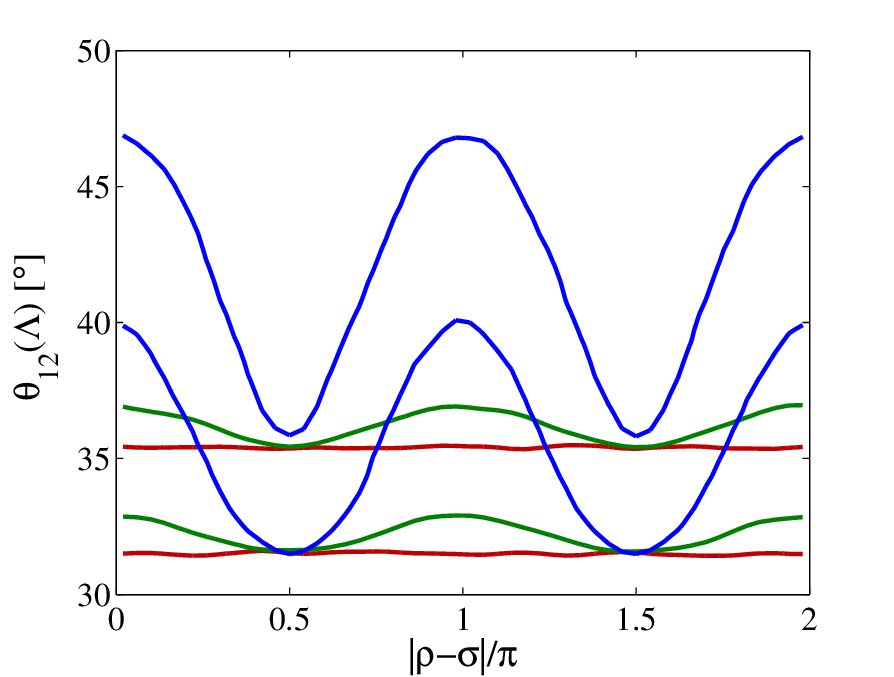}
\caption{\label{fig:fig-phase} The 90~\% posterior probability
regions for $\theta_{12}(\Lambda)$ as depending on the Majorana
CP-violating phase difference $\rho-\sigma$ in model~I (left panel)
and model~II (right panel). The bands between red, green, and blue
curves correspond to the allowed parameter space for
$m_1=0,~0.2~{\rm eV}$, and $0.5~{\rm eV}$, respectively.}
\vspace{-0.cm}
\end{center}
\end{figure}

Finally, since the RG running of $\theta_{12}$ is very sensitive to
the Majorana CP-violating phases, it is of interest to observe the
correlations between the CP-violating phases and the RG running of
the leptonic mixing angles. To this end, we show in
Fig.~\ref{fig:fig-phase} the correlations between $\theta_{12}$ at
the cutoff scale and the phase difference $\rho-\sigma$, with
$5\cdot 10^4$ samples per value of $\rho-\sigma$.
The bands in the plot indicate the favored parameter spaces (at
$90~\%$ posterior probability) for $m_1=0$ (red), $0.2~{\rm eV}$
(green), and $0.5~{\rm eV}$ (blue), respectively. As expected, no
visible RG running effects are observed for a hierarchical neutrino
spectrum. As for the degenerate neutrino spectrum, sizable RG
corrections may exist depending on the difference between two
Majorana CP-violating phases. A peak in $\theta_{12} (\Lambda)$
appears around $\rho-\sigma = n \pi$, while the RG corrections are
damped if $\rho-\sigma = (n+1)\pi/2$, with $n$ being an integer.

\section{Summary and conclusion}\label{sec:Summary}

In this work, we have studied the RG running of neutrino parameters
in two representative extra-dimensional models extended by an
effective neutrino mass operator. In particular, we have derived the
full set of RGEs for the Yukawa coupling matrices and the neutrino
mass operator. Both analytical and numerical analyzes of RG
corrections to the neutrino mixing angles and masses have been
performed according to our RGEs. We have found that, due to the
power-law behavior and a sizable enhancement factor, $\theta_{12}$
is the most sensitive mixing angle to the RG running, especially in
the nearly degenerate limit, \eg, $\theta_{23} = \pi/4$ can be
accommodated at high energy scales if $m_1 \simeq 0.5~{\rm eV}$.
In contrast, the RG corrections to $\theta_{23}$ and $\theta_{13}$
are negligible in both of the models. In addition, the RG running
effects on $\theta_{12}$ in model~I are generally larger than those in
model~II, due to the charged-lepton Yukawa couplings. Most
interestingly, $\theta_{12}$ does not decrease with increasing
energy scale, regardless of the choice of the neutrino mass hierarchy
and the CP-violating phases. Therefore, mixing patterns with small
$\theta_{12}$ at the cutoff scale are not compatible with low-energy
experiments, whereas the tri-bimaximal and bi-maximal patterns turn
out to be favorable, depending on the specific choice of model
parameters. We have also presented the connection between the
running $\theta_{12}$ and the Majorana CP-violating phases, which
indicates that the RG correction is damped if the difference between
the Majorana phases is close to $(2n+1)\pi /2$. We conclude that
quantum corrections should not be neglected in studies of
extra-dimensional neutrino mass models. Our results allow to carry
out an integrated investigation of fermion masses and flavor mixing
in the framework of extra dimensions.

In the current work, we have only considered theories with one spatial extra dimension. In the cases of models with more than one extra dimension, the RG corrections are generally more substantial. The conclusions of our study depend on the compactification scheme, and other choices of boundary conditions would lead to different phenomena, which are however beyond the scope of the current work.

\begin{acknowledgments}
This work was supported by the European
Commission Marie Curie Actions Framework Programme 7 Intra-European
Fellowship: Neutrino Evolution (M.B.), the Swedish Research
Council (Vetenskapsr{\aa}det), contract no.\ 621-2008-4210 (T.O.), as well as the ERC under the Starting Grant MANITOP
and the Deutsche Forschungsgemeinschaft in the Transregio 27
``Neutrinos and beyond -- weakly interacting particles in physics,
astrophysics and cosmology'' (H.Z.).
\end{acknowledgments}

\newpage
\appendix
\section{Full set of one-loop RGEs for the extra-dimensional models}
\label{sec:appendix}

The one-loop RGEs for the Yukawa coupling matrices $Y_f$ ($f=u,d,\ell$)
and the neutrino mass operator $\kappa$ can be expressed in a general
form as
\begin{eqnarray}
16\pi^2 \frac{{\rm d}Y_f}{{\rm d}\ln \mu} & = &  \beta^{\rm SM}_f +
\tilde\beta_f = \beta^{\rm SM}_f + \alpha_f Y_f + Y_f N_f \, ,  \\
16\pi^2 \frac{{\rm d}\kappa}{{\rm d}\ln \mu} &= &  \beta^{\rm
SM}_\kappa + \tilde \beta_\kappa =  \beta^{\rm SM}_\kappa +
\alpha_\kappa \kappa + \kappa N_\kappa +  N^T_\kappa \kappa \, ,
\end{eqnarray}
where the SM beta functions
are~\cite{Babu:1993qv,Chankowski:1993tx,Antusch:2001vn}
\begin{eqnarray}\label{eq:RGE-Y-SM}
\beta^{\rm SM}_u  & = & Y_u \left(\frac{3}{2} Y^\dagger_u Y_u -
\frac{3}{2} Y^\dagger_d Y_d - \frac{17}{12} g^2_1 - \frac{9}{4}
g^2_2 - 8g^2_3 +T\right) \, ,
\\
\beta^{\rm SM}_d & = & Y_d \left( -\frac{3}{2}  Y^\dagger_u Y_u +
\frac{3}{2} Y^\dagger_d Y_d - \frac{5}{12} g^2_1 - \frac{9}{4} g^2_2
- 8g^2_3 +T \right)\, ,
\\
\beta^{\rm SM}_\ell & = & Y_\ell \left( \frac{3}{2}  Y^\dagger_\ell
Y_\ell -
\frac{15}{4} g^2_1 - \frac{9}{4}g^2_2+T\right) \, , \\
\beta^{\rm SM}_\kappa & = & -\frac{3}{2} \kappa \left(
Y^\dagger_\ell Y_\ell \right)-\frac{3}{2}  \left( Y^\dagger_\ell
Y_\ell \right)^T \kappa  + \left(\lambda - 3 g^2_2 +2T\right) \kappa
\, ,
\end{eqnarray}
with $T= {\rm tr} \left(3 Y^\dagger_u Y_u + 3 Y^\dagger_d Y_d +
Y^\dagger_\ell Y_\ell \right)$. The contributions from KK
excitations are given by
\begin{eqnarray}
N_u  & = & \frac{3}{2} \eta s Y^\dagger_u Y_u - \frac{3}{2} \eta s
Y^\dagger_d Y_d \, ,
\\
N_d  & = & -\frac{3}{2} \eta s Y^\dagger_u Y_u + \frac{3}{2} \eta s
Y^\dagger_d Y_d \, ,
\\
N_\ell & = & \frac{3}{2} \eta s Y^\dagger_\ell Y_\ell \, , \\
N_\kappa & = & -\frac{3}{2} \eta s Y^\dagger_\ell Y_\ell \, ,
\end{eqnarray}
where $\eta=1$ for model~I and $\eta=2$ for model~II. Here, we
have defined the scale parameter $s=\lfloor \mu/\mu_0 \rfloor$. The flavor diagonal
coefficients $\alpha$'s for model~I read
\begin{eqnarray}\label{eq:alphas-II}
\alpha_u & = & \eta s \left(- \frac{101}{72} g^2_1 - \frac{15}{8}
g^2_2 - \frac{28}{3}g^2_3 + 2T \right)  \, , \\
\alpha_d & = & \eta s \left(-\frac{17}{72} g^2_1 - \frac{15}{8}
g^2_2 - \frac{28}{3} g^2_3 + 2T\right)  \, , \\
\alpha_\ell & = & \eta s \left(- \frac{33}{8} g^2_1 - \frac{15}{8}
g^2_2 + 2 T  \right) \, , \label{eq:alphal} \\
\alpha_\kappa & = & \eta s  \left( -\frac{1}{4} g^2_1
 - \frac{11}{4} g^2_2 + 4 T + \lambda  \right)
 \, ,
\end{eqnarray}
whereas for model~II, we have
\begin{eqnarray}\label{eq:alphas-I}
\alpha_u & = & \eta s \left(- \frac{17}{12} g^2_1 - \frac{9}{4}
g^2_2 - 8g^2_3 \right) \, , \\
\alpha_d & = & \eta s \left(- \frac{5}{12} g^2_1 - \frac{9}{4}
g^2_2 - 8g^2_3\right) \, , \\
\alpha_\ell & = & \eta s \left(- \frac{15}{4} g^2_1 - \frac{9}{4}
g^2_2\right)  \, , \\
\alpha_\kappa & = & \eta s  \left(\lambda - 3 g^2_2\right)
 \, .
\end{eqnarray}

For the sake of completeness, we also present the RGEs for the gauge
couplings
\begin{eqnarray}\label{eq:RGE-g}
16\pi^2 \frac{{\rm d}g_i}{{\rm d}\ln \mu} = \left(b^{\rm SM}_i +
\eta s \tilde b_i \right) g^3_i \, ,
\end{eqnarray}
where $(b^{\rm SM}_1,b^{\rm SM}_2,b^{\rm SM}_3) = (41/6,-19/6,-7)$,
while $(\tilde b_1,\tilde b_2,\tilde b_3) = (27/2,7/6,-5/2)$ in
model~I and $(\tilde b_1,\tilde b_2,\tilde b_3) = (1/6,-41/6,-21/2)$
in model~II. Furthermore, we also need the running of the Higgs
self-coupling,
\begin{eqnarray}\label{eq:RGE-lambda}
16\pi^2 \frac{{\rm d}\lambda}{{\rm d}\ln \mu} = \beta_\lambda^{\rm
SM} + \tilde \beta_\lambda \, ,
\end{eqnarray}
where the SM contribution reads~\cite{Cheng:1973nv}
\begin{eqnarray}
\beta_\lambda^{\rm SM} & = & 6 \lambda^2 - \lambda\left( 3 g^2_1 +9
g^2_2 \right) + \left( \frac{3}{2}g^4_1 + 3 g^2_1 g^2_2  +
\frac{9}{2}g^4_2 \right) \nonumber \\ &&+4 \lambda T -8 {\rm
tr}\left[3 \left(Y^\dagger_u Y_u\right)^2 + 3 \left(Y^\dagger_d Y_d
\right)^2+ \left(Y^\dagger_\ell Y_\ell\right)^2 \right] \, .
\end{eqnarray}
In addition, the extra-dimensional contributions are
\begin{eqnarray}
\tilde \beta_\lambda & = & 6 \eta s \lambda^2 - \eta s \lambda\left(
3 g^2_1 + 9 g^2_2 \right) + \eta s \left( 2 g^4_1 + 4 g^2_1
g^2_2 + 6 g^4_2 \right) \nonumber \\
&& + 8 \eta s \lambda T - 16 \eta s {\rm tr}\left[ 3
\left(Y^\dagger_u Y_u\right)^2 + 3 \left(Y^\dagger_d Y_d \right)^2 +
\left(Y^\dagger_\ell Y_\ell\right)^2 \right] \, ,
\end{eqnarray}
for model~I, and
\begin{eqnarray}
\tilde \beta_\lambda & = & 6 \eta s \lambda^2 - \eta s \lambda
\left( 3 g^2_1 +9g^2_2 \right) +  \eta s  \left( \frac{3}{2}g^4_1 +
3 g^2_1 g^2_2 + \frac{9}{2}g^4_2 \right) \, ,
\label{eq:beta_tilde_lambda}
\end{eqnarray}
for model~II.
Note that it has been pointed out that there may be a discrepancy between the factor $\tfrac{3}{2} g_1^4 + 3 g_1^2 g_2^2 + \tfrac{9}{2} g_2^4$ in Eq.~(\ref{eq:beta_tilde_lambda}) and the corresponding factor in Eq.~(A7) of Ref.~\cite{Liu:2012mea}. The factor in this work is consistent with the related factor in Eq.~(4) of Ref.~\cite{Bhattacharyya:2002nc}. In addition, the factor $\eta$ in Eq.~(\ref{eq:RGE-g}) is consistent with the findings in Refs.~\cite{Dienes:1998vg,Bhattacharyya:2002nc}.


\begin{thebibliography}{10}%
\makeatletter
\providecommand \@ifxundefined [1]{%
 \ifx #1\undefined \expandafter \@firstoftwo
 \else \expandafter \@secondoftwo
\fi
}%
\providecommand \@ifnum [1]{%
 \ifnum #1\expandafter \@firstoftwo
 \else \expandafter \@secondoftwo
\fi
}%
\providecommand \enquote [1]{``#1''}%
\providecommand \bibnamefont  [1]{#1}%
\providecommand \bibfnamefont [1]{#1}%
\providecommand \citenamefont [1]{#1}%
\providecommand\href[0]{\@sanitize\@href}%
\providecommand\@href[1]{\endgroup\@@startlink{#1}\endgroup\@@href}%
\providecommand\@@href[1]{#1\@@endlink}%
\providecommand \@sanitize [0]{\begingroup\catcode`\&12\catcode`\#12\relax}%
\@ifxundefined \pdfoutput {\@firstoftwo}{%
 \@ifnum{\z@=\pdfoutput}{\@firstoftwo}{\@secondoftwo}%
}{%
 \providecommand\@@startlink[1]{\leavevmode}%
 \providecommand\@@endlink[0]{}%
}{%
 \providecommand\@@startlink[1]{%
  \leavevmode
  \pdfstartlink
   attr{/Border[0 0 1 ]/H/I/C[0 1 1]}%
   user{/Subtype/Link/A<</Type/Action/S/URI/URI(#1)>>}%
  \relax
 }%
 \providecommand\@@endlink[0]{\pdfendlink}%
}%
\providecommand \url  [0]{\begingroup\@sanitize \@url }%
\providecommand \@url [1]{\endgroup\@href {#1}{\urlprefix}}%
\providecommand \urlprefix [0]{URL }%
\providecommand \Eprint[0]{\href }%
\@ifxundefined \urlstyle {%
  \providecommand \doi [1]{doi:\discretionary{}{}{}#1}%
}{%
  \providecommand \doi [0]{doi:\discretionary{}{}{}\begingroup
  \urlstyle{rm}\Url }%
}%
\providecommand \doibase [0]{http://dx.doi.org/}%
\providecommand \Doi[1]{\href{\doibase#1}}%
\providecommand \bibAnnote [3]{%
  \BibitemShut{#1}%
  \begin{quotation}\noindent
    \textsc{Key:}\ #2\\\textsc{Annotation:}\ #3%
  \end{quotation}%
}%
\providecommand \bibAnnoteFile [2]{%
  \IfFileExists{#2}{\bibAnnote {#1} {#2} {\input{#2}}}{}%
}%
\providecommand \typeout [0]{\immediate \write \m@ne }%
\providecommand \selectlanguage [0]{\@gobble}%
\providecommand \bibinfo [0]{\@secondoftwo}%
\providecommand \bibfield [0]{\@secondoftwo}%
\providecommand \translation [1]{[#1]}%
\providecommand \BibitemOpen[0]{}%
\providecommand \bibitemStop [0]{}%
\providecommand \bibitemNoStop [0]{.\EOS\space}%
\providecommand \EOS [0]{\spacefactor3000\relax}%
\providecommand \BibitemShut [1]{\csname bibitem#1\endcsname}%
\bibitem{Kaluza:1921tu}%
  \BibitemOpen
  \bibfield{author}{%
  \bibinfo {author} {\bibfnamefont{T.}~\bibnamefont{Kaluza}},\ }%
  \bibfield{journal}{%
  \bibinfo {journal} {Sitzungsber. Preuss. Akad. Wiss. Berlin (Math. Phys. )}\
  }%
  \textbf{\bibinfo {volume} {1921}},\ \bibinfo {pages} {966} (\bibinfo {year}
  {1921})%
  \bibAnnoteFile{NoStop}{Kaluza:1921tu}%
\bibitem{Klein:1926tv}%
  \BibitemOpen
  \bibfield{author}{%
  \bibinfo {author} {\bibfnamefont{O.}~\bibnamefont{Klein}},\ }%
  \bibfield{journal}{%
  \bibinfo {journal} {Z. Phys.}\ }%
  \textbf{\bibinfo {volume} {37}},\ \bibinfo {pages} {895} (\bibinfo {year}
  {1926})%
  \bibAnnoteFile{NoStop}{Klein:1926tv}%
\bibitem{Antoniadis:1998ig}%
  \BibitemOpen
  \bibfield{author}{%
  \bibinfo {author} {\bibfnamefont{I.}~\bibnamefont{Antoniadis}}, \bibinfo
  {author} {\bibfnamefont{N.}~\bibnamefont{Arkani-Hamed}}, \bibinfo {author}
  {\bibfnamefont{S.}~\bibnamefont{Dimopoulos}},\ and\ \bibinfo {author}
  {\bibfnamefont{G.}~\bibnamefont{Dvali}},\ }%
  \bibfield{journal}{%
  \Doi{10.1016/S0370-2693(98)00860-0}{\bibinfo {journal} {Phys.Lett.}}\ }%
  \textbf{\bibinfo {volume} {B436}},\ \bibinfo {pages} {257} (\bibinfo {year}
  {1998}),\ \Eprint{http://arxiv.org/abs/hep-ph/9804398}{arXiv:hep-ph/9804398
  [hep-ph]}%
  \bibAnnoteFile{NoStop}{Antoniadis:1998ig}%
\bibitem{ArkaniHamed:1998rs}%
  \BibitemOpen
  \bibfield{author}{%
  \bibinfo {author} {\bibfnamefont{N.}~\bibnamefont{Arkani-Hamed}}, \bibinfo
  {author} {\bibfnamefont{S.}~\bibnamefont{Dimopoulos}},\ and\ \bibinfo
  {author} {\bibfnamefont{G.~R.}\ \bibnamefont{Dvali}},\ }%
  \bibfield{journal}{%
  \bibinfo {journal} {Phys. Lett.}\ }%
  \textbf{\bibinfo {volume} {B429}},\ \bibinfo {pages} {263} (\bibinfo {year}
  {1998}),\ \Eprint{http://arxiv.org/abs/hep-ph/9803315}{hep-ph/9803315}%
  \bibAnnoteFile{NoStop}{ArkaniHamed:1998rs}%
\bibitem{ArkaniHamed:1998nn}%
  \BibitemOpen
  \bibfield{author}{%
  \bibinfo {author} {\bibfnamefont{N.}~\bibnamefont{Arkani-Hamed}}, \bibinfo
  {author} {\bibfnamefont{S.}~\bibnamefont{Dimopoulos}},\ and\ \bibinfo
  {author} {\bibfnamefont{G.~R.}\ \bibnamefont{Dvali}},\ }%
  \bibfield{journal}{%
  \bibinfo {journal} {Phys. Rev.}\ }%
  \textbf{\bibinfo {volume} {D59}},\ \bibinfo {pages} {086004} (\bibinfo {year}
  {1999}),\ \Eprint{http://arxiv.org/abs/hep-ph/9807344}{hep-ph/9807344}%
  \bibAnnoteFile{NoStop}{ArkaniHamed:1998nn}%
\bibitem{Randall:1999vf}%
  \BibitemOpen
  \bibfield{author}{%
  \bibinfo {author} {\bibfnamefont{L.}~\bibnamefont{Randall}}\ and\ \bibinfo
  {author} {\bibfnamefont{R.}~\bibnamefont{Sundrum}},\ }%
  \bibfield{journal}{%
  \bibinfo {journal} {Phys. Rev. Lett.}\ }%
  \textbf{\bibinfo {volume} {83}},\ \bibinfo {pages} {4690} (\bibinfo {year}
  {1999}),\ \Eprint{http://arxiv.org/abs/hep-th/9906064}{hep-th/9906064}%
  \bibAnnoteFile{NoStop}{Randall:1999vf}%
\bibitem{Randall:1999ee}%
  \BibitemOpen
  \bibfield{author}{%
  \bibinfo {author} {\bibfnamefont{L.}~\bibnamefont{Randall}}\ and\ \bibinfo
  {author} {\bibfnamefont{R.}~\bibnamefont{Sundrum}},\ }%
  \bibfield{journal}{%
  \bibinfo {journal} {Phys. Rev. Lett.}\ }%
  \textbf{\bibinfo {volume} {83}},\ \bibinfo {pages} {3370} (\bibinfo {year}
  {1999}),\ \Eprint{http://arxiv.org/abs/hep-ph/9905221}{hep-ph/9905221}%
  \bibAnnoteFile{NoStop}{Randall:1999ee}%
\bibitem{Appelquist:2000nn}%
  \BibitemOpen
  \bibfield{author}{%
  \bibinfo {author} {\bibfnamefont{T.}~\bibnamefont{Appelquist}}, \bibinfo
  {author} {\bibfnamefont{H.-C.}\ \bibnamefont{Cheng}},\ and\ \bibinfo {author}
  {\bibfnamefont{B.~A.}\ \bibnamefont{Dobrescu}},\ }%
  \bibfield{journal}{%
  \Doi{10.1103/PhysRevD.64.035002}{\bibinfo {journal} {Phys. Rev.}}\ }%
  \textbf{\bibinfo {volume} {D64}},\ \bibinfo {pages} {035002} (\bibinfo {year}
  {2001}),\ \Eprint{http://arxiv.org/abs/hep-ph/0012100}{arXiv:hep-ph/0012100}%
  \bibAnnoteFile{NoStop}{Appelquist:2000nn}%
\bibitem{Dienes:1998vg}%
  \BibitemOpen
  \bibfield{author}{%
  \bibinfo {author} {\bibfnamefont{K.~R.}\ \bibnamefont{Dienes}}, \bibinfo
  {author} {\bibfnamefont{E.}~\bibnamefont{Dudas}},\ and\ \bibinfo {author}
  {\bibfnamefont{T.}~\bibnamefont{Gherghetta}},\ }%
  \bibfield{journal}{%
  \Doi{10.1016/S0550-3213(98)00669-5}{\bibinfo {journal} {Nucl. Phys.}}\ }%
  \textbf{\bibinfo {volume} {B537}},\ \bibinfo {pages} {47} (\bibinfo {year}
  {1999}),\ \Eprint{http://arxiv.org/abs/hep-ph/9806292}{arXiv:hep-ph/9806292}%
  \bibAnnoteFile{NoStop}{Dienes:1998vg}%
\bibitem{Minkowski:1977sc}%
  \BibitemOpen
  \bibfield{author}{%
  \bibinfo {author} {\bibfnamefont{P.}~\bibnamefont{Minkowski}},\ }%
  \bibfield{journal}{%
  \bibinfo {journal} {Phys. Lett.}\ }%
  \textbf{\bibinfo {volume} {B67}},\ \bibinfo {pages} {421} (\bibinfo {year}
  {1977})%
  \bibAnnoteFile{NoStop}{Minkowski:1977sc}%
\bibitem{Yanagida:1979as}%
  \BibitemOpen
  \bibfield{author}{%
  \bibinfo {author} {\bibfnamefont{T.}~\bibnamefont{Yanagida}},\ }%
  in\ \emph{\bibinfo {booktitle} {Proc. Workshop on the baryon number of the
  Universe and unified theories}},\ \bibinfo {editor} {edited by\ \bibinfo
  {editor} {\bibfnamefont{O.}~\bibnamefont{Sawada}}\ and\ \bibinfo {editor}
  {\bibfnamefont{A.}~\bibnamefont{Sugamoto}}}\ (\bibinfo {year} {1979})\
  p.~\bibinfo {pages} {95}%
  \bibAnnoteFile{NoStop}{Yanagida:1979as}%
\bibitem{Mohapatra:1979ia}%
  \BibitemOpen
  \bibfield{author}{%
  \bibinfo {author} {\bibfnamefont{R.~N.}\ \bibnamefont{Mohapatra}}\ and\
  \bibinfo {author} {\bibfnamefont{G.}~\bibnamefont{Senjanovi{\'c}}},\ }%
  \bibfield{journal}{%
  \bibinfo {journal} {Phys. Rev. Lett.}\ }%
  \textbf{\bibinfo {volume} {44}},\ \bibinfo {pages} {912} (\bibinfo {year}
  {1980})%
  \bibAnnoteFile{NoStop}{Mohapatra:1979ia}%
\bibitem{GellMann:1980vs}%
  \BibitemOpen
  \bibfield{author}{%
  \bibinfo {author} {\bibfnamefont{M.}~\bibnamefont{Gell-Mann}}, \bibinfo
  {author} {\bibfnamefont{P.}~\bibnamefont{Ramond}},\ and\ \bibinfo {author}
  {\bibfnamefont{R.}~\bibnamefont{Slansky}},\ }%
  in\ \emph{\bibinfo {booktitle} {Supergravity}},\ \bibinfo {editor} {edited
  by\ \bibinfo {editor} {\bibfnamefont{P.}~\bibnamefont{van Nieuwenhuizen}}\
  and\ \bibinfo {editor} {\bibfnamefont{D.}~\bibnamefont{Freedman}}}\ (\bibinfo
  {year} {1979})\ p.\ \bibinfo {pages} {315}%
  \bibAnnoteFile{NoStop}{GellMann:1980vs}%
\bibitem{Frere:2003hn}%
  \BibitemOpen
  \bibfield{author}{%
  \bibinfo {author} {\bibfnamefont{J.~M.}\ \bibnamefont{Frere}}, \bibinfo
  {author} {\bibfnamefont{G.}~\bibnamefont{Moreau}},\ and\ \bibinfo {author}
  {\bibfnamefont{E.}~\bibnamefont{Nezri}},\ }%
  \bibfield{journal}{%
  \Doi{10.1103/PhysRevD.69.033003}{\bibinfo {journal} {Phys. Rev.}}\ }%
  \textbf{\bibinfo {volume} {D69}},\ \bibinfo {pages} {033003} (\bibinfo {year}
  {2004}),\ \Eprint{http://arxiv.org/abs/hep-ph/0309218}{arXiv:hep-ph/0309218}%
  \bibAnnoteFile{NoStop}{Frere:2003hn}%
\bibitem{Haba:2009sd}%
  \BibitemOpen
  \bibfield{author}{%
  \bibinfo {author} {\bibfnamefont{N.}~\bibnamefont{Haba}}, \bibinfo {author}
  {\bibfnamefont{S.}~\bibnamefont{Matsumoto}},\ and\ \bibinfo {author}
  {\bibfnamefont{K.}~\bibnamefont{Yoshioka}},\ }%
  \bibfield{journal}{%
  \Doi{10.1016/j.physletb.2009.05.042}{\bibinfo {journal} {Phys. Lett.}}\ }%
  \textbf{\bibinfo {volume} {B677}},\ \bibinfo {pages} {291} (\bibinfo {year}
  {2009}),\ \Eprint{http://arxiv.org/abs/0901.4596}{arXiv:0901.4596 [hep-ph]}%
  \bibAnnoteFile{NoStop}{Haba:2009sd}%
\bibitem{Blennow:2010zu}%
  \BibitemOpen
  \bibfield{author}{%
  \bibinfo {author} {\bibfnamefont{M.}~\bibnamefont{Blennow}}, \bibinfo
  {author} {\bibfnamefont{H.}~\bibnamefont{Melb{\'e}us}}, \bibinfo {author}
  {\bibfnamefont{T.}~\bibnamefont{Ohlsson}},\ and\ \bibinfo {author}
  {\bibfnamefont{H.}~\bibnamefont{Zhang}},\ }%
  \bibfield{journal}{%
  \Doi{10.1103/PhysRevD.82.045023}{\bibinfo {journal} {Phys. Rev.}}\ }%
  \textbf{\bibinfo {volume} {D82}},\ \bibinfo {pages} {045023} (\bibinfo {year}
  {2010}),\ \Eprint{http://arxiv.org/abs/1003.0669}{arXiv:1003.0669 [hep-ph]}%
  \bibAnnoteFile{NoStop}{Blennow:2010zu}%
\bibitem{Saito:2010xj}%
  \BibitemOpen
  \bibfield{author}{%
  \bibinfo {author} {\bibfnamefont{T.}~\bibnamefont{Saito}}, \bibinfo {author}
  {\bibfnamefont{M.}~\bibnamefont{Asano}}, \bibinfo {author}
  {\bibfnamefont{K.}~\bibnamefont{Fujii}}, \bibinfo {author}
  {\bibfnamefont{N.}~\bibnamefont{Haba}}, \bibinfo {author}
  {\bibfnamefont{S.}~\bibnamefont{Matsumoto}}, \emph{et~al.},\ }%
  \bibfield{journal}{%
  \Doi{10.1103/PhysRevD.82.093004}{\bibinfo {journal} {Phys. Rev.}}\ }%
  \textbf{\bibinfo {volume} {D82}},\ \bibinfo {pages} {093004} (\bibinfo {year}
  {2010}),\ \Eprint{http://arxiv.org/abs/1008.2257}{arXiv:1008.2257 [hep-ph]}%
  \bibAnnoteFile{NoStop}{Saito:2010xj}%
\bibitem{Chankowski:1993tx}%
  \BibitemOpen
  \bibfield{author}{%
  \bibinfo {author} {\bibfnamefont{P.~H.}\ \bibnamefont{Chankowski}}\ and\
  \bibinfo {author} {\bibfnamefont{Z.}~\bibnamefont{Pluciennik}},\ }%
  \bibfield{journal}{%
  \Doi{10.1016/0370-2693(93)90330-K}{\bibinfo {journal} {Phys. Lett.}}\ }%
  \textbf{\bibinfo {volume} {B316}},\ \bibinfo {pages} {312} (\bibinfo {year}
  {1993}),\ \Eprint{http://arxiv.org/abs/hep-ph/9306333}{arXiv:hep-ph/9306333}%
  \bibAnnoteFile{NoStop}{Chankowski:1993tx}%
\bibitem{Babu:1993qv}%
  \BibitemOpen
  \bibfield{author}{%
  \bibinfo {author} {\bibfnamefont{K.~S.}\ \bibnamefont{Babu}}, \bibinfo
  {author} {\bibfnamefont{C.~N.}\ \bibnamefont{Leung}},\ and\ \bibinfo {author}
  {\bibfnamefont{J.~T.}\ \bibnamefont{Pantaleone}},\ }%
  \bibfield{journal}{%
  \Doi{10.1016/0370-2693(93)90801-N}{\bibinfo {journal} {Phys. Lett.}}\ }%
  \textbf{\bibinfo {volume} {B319}},\ \bibinfo {pages} {191} (\bibinfo {year}
  {1993}),\ \Eprint{http://arxiv.org/abs/hep-ph/9309223}{arXiv:hep-ph/9309223}%
  \bibAnnoteFile{NoStop}{Babu:1993qv}%
\bibitem{Antusch:2001ck}%
  \BibitemOpen
  \bibfield{author}{%
  \bibinfo {author} {\bibfnamefont{S.}~\bibnamefont{Antusch}}, \bibinfo
  {author} {\bibfnamefont{M.}~\bibnamefont{Drees}}, \bibinfo {author}
  {\bibfnamefont{J.}~\bibnamefont{Kersten}}, \bibinfo {author}
  {\bibfnamefont{M.}~\bibnamefont{Lindner}},\ and\ \bibinfo {author}
  {\bibfnamefont{M.}~\bibnamefont{Ratz}},\ }%
  \bibfield{journal}{%
  \Doi{10.1016/S0370-2693(01)01127-3}{\bibinfo {journal} {Phys. Lett.}}\ }%
  \textbf{\bibinfo {volume} {B519}},\ \bibinfo {pages} {238} (\bibinfo {year}
  {2001}),\ \Eprint{http://arxiv.org/abs/hep-ph/0108005}{arXiv:hep-ph/0108005}%
  \bibAnnoteFile{NoStop}{Antusch:2001ck}%
\bibitem{Antusch:2001vn}%
  \BibitemOpen
  \bibfield{author}{%
  \bibinfo {author} {\bibfnamefont{S.}~\bibnamefont{Antusch}}, \bibinfo
  {author} {\bibfnamefont{M.}~\bibnamefont{Drees}}, \bibinfo {author}
  {\bibfnamefont{J.}~\bibnamefont{Kersten}}, \bibinfo {author}
  {\bibfnamefont{M.}~\bibnamefont{Lindner}},\ and\ \bibinfo {author}
  {\bibfnamefont{M.}~\bibnamefont{Ratz}},\ }%
  \bibfield{journal}{%
  \Doi{10.1016/S0370-2693(01)01414-9}{\bibinfo {journal} {Phys. Lett.}}\ }%
  \textbf{\bibinfo {volume} {B525}},\ \bibinfo {pages} {130} (\bibinfo {year}
  {2002}),\ \Eprint{http://arxiv.org/abs/hep-ph/0110366}{arXiv:hep-ph/0110366}%
  \bibAnnoteFile{NoStop}{Antusch:2001vn}%
\bibitem{Chao:2006ye}%
  \BibitemOpen
  \bibfield{author}{%
  \bibinfo {author} {\bibfnamefont{W.}~\bibnamefont{Chao}}\ and\ \bibinfo
  {author} {\bibfnamefont{H.}~\bibnamefont{Zhang}},\ }%
  \bibfield{journal}{%
  \Doi{10.1103/PhysRevD.75.033003}{\bibinfo {journal} {Phys. Rev.}}\ }%
  \textbf{\bibinfo {volume} {D75}},\ \bibinfo {pages} {033003} (\bibinfo {year}
  {2007}),\ \Eprint{http://arxiv.org/abs/hep-ph/0611323}{arXiv:hep-ph/0611323}%
  \bibAnnoteFile{NoStop}{Chao:2006ye}%
\bibitem{Schmidt:2007nq}%
  \BibitemOpen
  \bibfield{author}{%
  \bibinfo {author} {\bibfnamefont{M.~A.}\ \bibnamefont{Schmidt}},\ }%
  \bibfield{journal}{%
  \Doi{10.1103/PhysRevD.76.073010}{\bibinfo {journal} {Phys. Rev.}}\ }%
  \textbf{\bibinfo {volume} {D76}},\ \bibinfo {pages} {073010} (\bibinfo {year}
  {2007}),\ \Eprint{http://arxiv.org/abs/0705.3841}{arXiv:0705.3841 [hep-ph]}%
  \bibAnnoteFile{NoStop}{Schmidt:2007nq}%
\bibitem{Chakrabortty:2008zh}%
  \BibitemOpen
  \bibfield{author}{%
  \bibinfo {author} {\bibfnamefont{J.}~\bibnamefont{Chakrabortty}}, \bibinfo
  {author} {\bibfnamefont{A.}~\bibnamefont{Dighe}}, \bibinfo {author}
  {\bibfnamefont{S.}~\bibnamefont{Goswami}},\ and\ \bibinfo {author}
  {\bibfnamefont{S.}~\bibnamefont{Ray}},\ }%
  \bibfield{journal}{%
  \Doi{10.1016/j.nuclphysb.2009.05.016}{\bibinfo {journal} {Nucl. Phys.}}\ }%
  \textbf{\bibinfo {volume} {B820}},\ \bibinfo {pages} {116} (\bibinfo {year}
  {2009}),\ \Eprint{http://arxiv.org/abs/0812.2776}{arXiv:0812.2776 [hep-ph]}%
  \bibAnnoteFile{NoStop}{Chakrabortty:2008zh}%
\bibitem{Bergstrom:2010qb}%
  \BibitemOpen
  \bibfield{author}{%
  \bibinfo {author} {\bibfnamefont{J.}~\bibnamefont{Bergstr{\"o}m}}, \bibinfo
  {author} {\bibfnamefont{M.}~\bibnamefont{Malinsk{\'y}}}, \bibinfo {author}
  {\bibfnamefont{T.}~\bibnamefont{Ohlsson}},\ and\ \bibinfo {author}
  {\bibfnamefont{H.}~\bibnamefont{Zhang}},\ }%
  \bibfield{journal}{%
  \Doi{10.1103/PhysRevD.81.116006}{\bibinfo {journal} {Phys. Rev.}}\ }%
  \textbf{\bibinfo {volume} {D81}},\ \bibinfo {pages} {116006} (\bibinfo {year}
  {2010}),\ \Eprint{http://arxiv.org/abs/1004.4628}{arXiv:1004.4628 [hep-ph]}%
  \bibAnnoteFile{NoStop}{Bergstrom:2010qb}%
\bibitem{Hooper:2007qk}%
  \BibitemOpen
  \bibfield{author}{%
  \bibinfo {author} {\bibfnamefont{D.}~\bibnamefont{Hooper}}\ and\ \bibinfo
  {author} {\bibfnamefont{S.}~\bibnamefont{Profumo}},\ }%
  \bibfield{journal}{%
  \Doi{10.1016/j.physrep.2007.09.003}{\bibinfo {journal} {Phys. Rept.}}\ }%
  \textbf{\bibinfo {volume} {453}},\ \bibinfo {pages} {29} (\bibinfo {year}
  {2007}),\ \Eprint{http://arxiv.org/abs/hep-ph/0701197}{arXiv:hep-ph/0701197
  [hep-ph]}%
  \bibAnnoteFile{NoStop}{Hooper:2007qk}%
\bibitem{Bhattacharyya:2006ym}%
  \BibitemOpen
  \bibfield{author}{%
  \bibinfo {author} {\bibfnamefont{G.}~\bibnamefont{Bhattacharyya}}, \bibinfo
  {author} {\bibfnamefont{A.}~\bibnamefont{Datta}}, \bibinfo {author}
  {\bibfnamefont{S.~K.}\ \bibnamefont{Majee}},\ and\ \bibinfo {author}
  {\bibfnamefont{A.}~\bibnamefont{Raychaudhuri}},\ }%
  \bibfield{journal}{%
  \Doi{10.1016/j.nuclphysb.2006.10.018}{\bibinfo {journal} {Nucl. Phys.}}\ }%
  \textbf{\bibinfo {volume} {B760}},\ \bibinfo {pages} {117} (\bibinfo {year}
  {2007}),\ \Eprint{http://arxiv.org/abs/hep-ph/0608208}{arXiv:hep-ph/0608208}%
  \bibAnnoteFile{NoStop}{Bhattacharyya:2006ym}%
\bibitem{Cornell:2010sz}%
  \BibitemOpen
  \bibfield{author}{%
  \bibinfo {author} {\bibfnamefont{A.~S.}\ \bibnamefont{Cornell}}\ and\
  \bibinfo {author} {\bibfnamefont{L.-X.}\ \bibnamefont{Liu}}}%
   (\bibinfo {year} {2010}),\
  \Eprint{http://arxiv.org/abs/1010.5522}{arXiv:1010.5522 [hep-ph]}%
  \bibAnnoteFile{NoStop}{Cornell:2010sz}%
\bibitem{Antoniadis:1990ew}%
  \BibitemOpen
  \bibfield{author}{%
  \bibinfo {author} {\bibfnamefont{I.}~\bibnamefont{Antoniadis}},\ }%
  \bibfield{journal}{%
  \Doi{10.1016/0370-2693(90)90617-F}{\bibinfo {journal} {Phys. Lett.}}\ }%
  \textbf{\bibinfo {volume} {B246}},\ \bibinfo {pages} {377} (\bibinfo {year}
  {1990})%
  \bibAnnoteFile{NoStop}{Antoniadis:1990ew}%
\bibitem{Dienes:1998vh}%
  \BibitemOpen
  \bibfield{author}{%
  \bibinfo {author} {\bibfnamefont{K.~R.}\ \bibnamefont{Dienes}}, \bibinfo
  {author} {\bibfnamefont{E.}~\bibnamefont{Dudas}},\ and\ \bibinfo {author}
  {\bibfnamefont{T.}~\bibnamefont{Gherghetta}},\ }%
  \bibfield{journal}{%
  \Doi{10.1016/S0370-2693(98)00977-0}{\bibinfo {journal} {Phys. Lett.}}\ }%
  \textbf{\bibinfo {volume} {B436}},\ \bibinfo {pages} {55} (\bibinfo {year}
  {1998}),\ \Eprint{http://arxiv.org/abs/hep-ph/9803466}{arXiv:hep-ph/9803466
  [hep-ph]}%
  \bibAnnoteFile{NoStop}{Dienes:1998vh}%
\bibitem{Abel:1998wa}%
  \BibitemOpen
  \bibfield{author}{%
  \bibinfo {author} {\bibfnamefont{S.~A.}\ \bibnamefont{Abel}}\ and\ \bibinfo
  {author} {\bibfnamefont{S.~F.}\ \bibnamefont{King}},\ }%
  \bibfield{journal}{%
  \Doi{10.1103/PhysRevD.59.095010}{\bibinfo {journal} {Phys. Rev.}}\ }%
  \textbf{\bibinfo {volume} {D59}},\ \bibinfo {pages} {095010} (\bibinfo {year}
  {1999}),\ \Eprint{http://arxiv.org/abs/hep-ph/9809467}{arXiv:hep-ph/9809467}%
  \bibAnnoteFile{NoStop}{Abel:1998wa}%
\bibitem{Bhattacharyya:2002nc}%
  \BibitemOpen
  \bibfield{author}{%
  \bibinfo {author} {\bibfnamefont{G.}~\bibnamefont{Bhattacharyya}}, \bibinfo
  {author} {\bibfnamefont{S.}~\bibnamefont{Goswami}},\ and\ \bibinfo {author}
  {\bibfnamefont{A.}~\bibnamefont{Raychaudhuri}},\ }%
  \bibfield{journal}{%
  \Doi{10.1103/PhysRevD.66.033008}{\bibinfo {journal} {Phys. Rev.}}\ }%
  \textbf{\bibinfo {volume} {D66}},\ \bibinfo {pages} {033008} (\bibinfo {year}
  {2002}),\ \Eprint{http://arxiv.org/abs/hep-ph/0202147}{arXiv:hep-ph/0202147}%
  \bibAnnoteFile{NoStop}{Bhattacharyya:2002nc}%
\bibitem{Rizzo:1999br}%
  \BibitemOpen
  \bibfield{author}{%
  \bibinfo {author} {\bibfnamefont{T.~G.}\ \bibnamefont{Rizzo}}\ and\ \bibinfo
  {author} {\bibfnamefont{J.~D.}\ \bibnamefont{Wells}},\ }%
  \bibfield{journal}{%
  \Doi{10.1103/PhysRevD.61.016007}{\bibinfo {journal} {Phys. Rev.}}\ }%
  \textbf{\bibinfo {volume} {D61}},\ \bibinfo {pages} {016007} (\bibinfo {year}
  {2000}),\ \Eprint{http://arxiv.org/abs/hep-ph/9906234}{arXiv:hep-ph/9906234}%
  \bibAnnoteFile{NoStop}{Rizzo:1999br}%
\bibitem{Maki:1962mu}%
  \BibitemOpen
  \bibfield{author}{%
  \bibinfo {author} {\bibfnamefont{Z.}~\bibnamefont{Maki}}, \bibinfo {author}
  {\bibfnamefont{M.}~\bibnamefont{Nakagawa}},\ and\ \bibinfo {author}
  {\bibfnamefont{S.}~\bibnamefont{Sakata}},\ }%
  \bibfield{journal}{%
  \Doi{10.1143/PTP.28.870}{\bibinfo {journal} {Prog. Theor. Phys.}}\ }%
  \textbf{\bibinfo {volume} {28}},\ \bibinfo {pages} {870} (\bibinfo {year}
  {1962})%
  \bibAnnoteFile{NoStop}{Maki:1962mu}%
\bibitem{Pontecorvo:1967fh}%
  \BibitemOpen
  \bibfield{author}{%
  \bibinfo {author} {\bibfnamefont{B.}~\bibnamefont{Pontecorvo}},\ }%
  \bibfield{journal}{%
  \bibinfo {journal} {Sov. Phys. JETP}\ }%
  \textbf{\bibinfo {volume} {26}},\ \bibinfo {pages} {984} (\bibinfo {year}
  {1968})%
  \bibAnnoteFile{NoStop}{Pontecorvo:1967fh}%
\bibitem{Antusch:2005gp}%
  \BibitemOpen
  \bibfield{author}{%
  \bibinfo {author} {\bibfnamefont{S.}~\bibnamefont{Antusch}}, \bibinfo
  {author} {\bibfnamefont{J.}~\bibnamefont{Kersten}}, \bibinfo {author}
  {\bibfnamefont{M.}~\bibnamefont{Lindner}}, \bibinfo {author}
  {\bibfnamefont{M.}~\bibnamefont{Ratz}},\ and\ \bibinfo {author}
  {\bibfnamefont{M.~A.}\ \bibnamefont{Schmidt}},\ }%
  \bibfield{journal}{%
  \Doi{10.1088/1126-6708/2005/03/024}{\bibinfo {journal} {JHEP}}\ }%
  \textbf{\bibinfo {volume} {03}},\ \bibinfo {pages} {024} (\bibinfo {year}
  {2005}),\ \Eprint{http://arxiv.org/abs/hep-ph/0501272}{arXiv:hep-ph/0501272}%
  \bibAnnoteFile{NoStop}{Antusch:2005gp}%
\bibitem{Mei:2005qp}%
  \BibitemOpen
  \bibfield{author}{%
  \bibinfo {author} {\bibfnamefont{J.-w.}\ \bibnamefont{Mei}},\ }%
  \bibfield{journal}{%
  \Doi{10.1103/PhysRevD.71.073012}{\bibinfo {journal} {Phys. Rev.}}\ }%
  \textbf{\bibinfo {volume} {D71}},\ \bibinfo {pages} {073012} (\bibinfo {year}
  {2005}),\ \Eprint{http://arxiv.org/abs/hep-ph/0502015}{arXiv:hep-ph/0502015}%
  \bibAnnoteFile{NoStop}{Mei:2005qp}%
\bibitem{Harrison:2002er}%
  \BibitemOpen
  \bibfield{author}{%
  \bibinfo {author} {\bibfnamefont{P.~F.}\ \bibnamefont{Harrison}}, \bibinfo
  {author} {\bibfnamefont{D.~H.}\ \bibnamefont{Perkins}},\ and\ \bibinfo
  {author} {\bibfnamefont{W.~G.}\ \bibnamefont{Scott}},\ }%
  \bibfield{journal}{%
  \Doi{10.1016/S0370-2693(02)01336-9}{\bibinfo {journal} {Phys. Lett.}}\ }%
  \textbf{\bibinfo {volume} {B530}},\ \bibinfo {pages} {167} (\bibinfo {year}
  {2002}),\ \Eprint{http://arxiv.org/abs/hep-ph/0202074}{arXiv:hep-ph/0202074}%
  \bibAnnoteFile{NoStop}{Harrison:2002er}%
\bibitem{Harrison:2002kp}%
  \BibitemOpen
  \bibfield{author}{%
  \bibinfo {author} {\bibfnamefont{P.~F.}\ \bibnamefont{Harrison}}\ and\
  \bibinfo {author} {\bibfnamefont{W.~G.}\ \bibnamefont{Scott}},\ }%
  \bibfield{journal}{%
  \Doi{10.1016/S0370-2693(02)01753-7}{\bibinfo {journal} {Phys. Lett.}}\ }%
  \textbf{\bibinfo {volume} {B535}},\ \bibinfo {pages} {163} (\bibinfo {year}
  {2002}),\ \Eprint{http://arxiv.org/abs/hep-ph/0203209}{arXiv:hep-ph/0203209}%
  \bibAnnoteFile{NoStop}{Harrison:2002kp}%
\bibitem{Xing:2002sw}%
  \BibitemOpen
  \bibfield{author}{%
  \bibinfo {author} {\bibfnamefont{Z.-z.}\ \bibnamefont{Xing}},\ }%
  \bibfield{journal}{%
  \Doi{10.1016/S0370-2693(02)01649-0}{\bibinfo {journal} {Phys. Lett.}}\ }%
  \textbf{\bibinfo {volume} {B533}},\ \bibinfo {pages} {85} (\bibinfo {year}
  {2002}),\ \Eprint{http://arxiv.org/abs/hep-ph/0204049}{arXiv:hep-ph/0204049}%
  \bibAnnoteFile{NoStop}{Xing:2002sw}%
\bibitem{Schwetz:2008er}%
  \BibitemOpen
  \bibfield{author}{%
  \bibinfo {author} {\bibfnamefont{T.}~\bibnamefont{Schwetz}}, \bibinfo
  {author} {\bibfnamefont{M.~A.}\ \bibnamefont{T{\'o}rtola}},\ and\ \bibinfo
  {author} {\bibfnamefont{J.~W.~F.}\ \bibnamefont{Valle}},\ }%
  \bibfield{journal}{%
  \Doi{10.1088/1367-2630/10/11/113011}{\bibinfo {journal} {New J. Phys.}}\ }%
  \textbf{\bibinfo {volume} {10}},\ \bibinfo {pages} {113011} (\bibinfo {year}
  {2008}),\ \Eprint{http://arxiv.org/abs/0808.2016}{arXiv:0808.2016 [hep-ph]}%
  \bibAnnoteFile{NoStop}{Schwetz:2008er}%
\bibitem{Xing:2007fb}%
  \BibitemOpen
  \bibfield{author}{%
  \bibinfo {author} {\bibfnamefont{Z.-z.}\ \bibnamefont{Xing}}, \bibinfo
  {author} {\bibfnamefont{H.}~\bibnamefont{Zhang}},\ and\ \bibinfo {author}
  {\bibfnamefont{S.}~\bibnamefont{Zhou}},\ }%
  \bibfield{journal}{%
  \Doi{10.1103/PhysRevD.77.113016}{\bibinfo {journal} {Phys. Rev.}}\ }%
  \textbf{\bibinfo {volume} {D77}},\ \bibinfo {pages} {113016} (\bibinfo {year}
  {2008}),\ \Eprint{http://arxiv.org/abs/0712.1419}{arXiv:0712.1419 [hep-ph]}%
  \bibAnnoteFile{NoStop}{Xing:2007fb}%
\bibitem{Lobashev:2003kt}%
  \BibitemOpen
  \bibfield{author}{%
  \bibinfo {author} {\bibfnamefont{V.}~\bibnamefont{Lobashev}},\ }%
  \bibfield{journal}{%
  \Doi{10.1016/S0375-9474(03)00985-0}{\bibinfo {journal} {Nucl. Phys.}}\ }%
  \textbf{\bibinfo {volume} {A719}},\ \bibinfo {pages} {153} (\bibinfo {year}
  {2003})%
  \bibAnnoteFile{NoStop}{Lobashev:2003kt}%
\bibitem{Eitel:2005hg}%
  \BibitemOpen
  \bibfield{author}{%
  \bibinfo {author} {\bibfnamefont{K.}~\bibnamefont{Eitel}},\ }%
  \bibfield{journal}{%
  \Doi{10.1016/j.nuclphysbps.2005.01.105}{\bibinfo {journal} {Nucl. Phys. Proc.
  Suppl.}}\ }%
  \textbf{\bibinfo {volume} {143}},\ \bibinfo {pages} {197} (\bibinfo {year}
  {2005})%
  \bibAnnoteFile{NoStop}{Eitel:2005hg}%
\bibitem{Blennow:2009pk}%
  \BibitemOpen
  \bibfield{author}{%
  \bibinfo {author} {\bibfnamefont{M.}~\bibnamefont{Blennow}}\ and\ \bibinfo
  {author} {\bibfnamefont{E.}~\bibnamefont{Fernandez-Martinez}},\ }%
  \bibfield{journal}{%
  \Doi{10.1016/j.cpc.2009.09.014}{\bibinfo {journal} {Comput. Phys. Commun.}}\
  }%
  \textbf{\bibinfo {volume} {181}},\ \bibinfo {pages} {227} (\bibinfo {year}
  {2010}),\ \Eprint{http://arxiv.org/abs/0903.3985}{arXiv:0903.3985 [hep-ph]}%
  \bibAnnoteFile{NoStop}{Blennow:2009pk}%
\bibitem{Cheng:1973nv}%
  \BibitemOpen
  \bibfield{author}{%
  \bibinfo {author} {\bibfnamefont{T.~P.}\ \bibnamefont{Cheng}}, \bibinfo
  {author} {\bibfnamefont{E.}~\bibnamefont{Eichten}},\ and\ \bibinfo {author}
  {\bibfnamefont{L.-F.}\ \bibnamefont{Li}},\ }%
  \bibfield{journal}{%
  \Doi{10.1103/PhysRevD.9.2259}{\bibinfo {journal} {Phys. Rev.}}\ }%
  \textbf{\bibinfo {volume} {D9}},\ \bibinfo {pages} {2259} (\bibinfo {year}
  {1974})%
  \bibAnnoteFile{NoStop}{Cheng:1973nv}%
\bibitem{Liu:2012mea}%
  \BibitemOpen
  \bibfield{author}{%
  \bibinfo {author} {\bibfnamefont{L.-X.}\ \bibnamefont{Liu}},\ and\ \bibinfo {author}
  {\bibfnamefont{A.~S.}\ \bibnamefont{Cornell}},\ }%
  \bibfield{journal}{%
  \Doi{10.1103/PhysRevD.86.056002}{\bibinfo {journal} {Phys. Rev.}}\ }%
  \textbf{\bibinfo {volume} {D86}},\ \bibinfo {pages} {056002} (\bibinfo {year}
  {2012}),\ \Eprint{http://arxiv.org/abs/1204.0532}{arXiv:1204.0532 [hep-ph]}%
  \bibAnnoteFile{NoStop}{Liu:2012mea}%
\end{thebibliography}
\end{document}